\documentclass[%
reprint,
superscriptaddress,
showpacs,
amsmath,amssymb,
aps,
prx,
longbibliography,
]{revtex4-1}

\usepackage{psfrag,graphicx,epsfig,color}
\usepackage{dcolumn}
\usepackage{bm}
\usepackage{natbib}
\usepackage[usenames,dvipsnames,svgnames,table]{xcolor}
\usepackage{subfigure}
\usepackage{rotating}
\usepackage[nice]{nicefrac}

\def\newblock{}

\def\re {{R_\lambda}}
\definecolor{mygreen}{rgb}{0,0.7,0.}

\def\AP#1{{\color{blue}{#1}}}

\begin{document}

\title{Extreme velocity gradients in turbulent flows}

\author{Dhawal Buaria }
\email[]{dhawal.buaria@ds.mpg.de}
\affiliation{Max Planck Institute for Dynamics and Self-Organization, 37077 G\"ottingen, Germany}
\author{Alain Pumir}
\affiliation{Laboratoire de Physique, Ecole Normale Sup\'erieure de Lyon, Universit\'e de Lyon 1 and Centre National de la Recherche Scientifique, 69007 Lyon, France}
\affiliation{Max Planck Institute for Dynamics and Self-Organization, 37077 G\"ottingen, Germany}

\author{Eberhard Bodenschatz}
\affiliation{Max Planck Institute for Dynamics and Self-Organization, 37077 G\"ottingen, Germany}
\affiliation{Institute for Nonlinear Dynamics, University of G\"ottingen, 37077 G\"ottingen, Germany}
\affiliation{Max Planck Center Twente, 7500 AE Enschede, The Netherlands} 
\affiliation{Laboratory of Atomic and Solid-State Physics and Sibley School of Mechanical and Aerospace Engineering, Cornell University, Ithaca, New York 14853, USA}

\author{P. K. Yeung }
\affiliation{Schools of Aerospace Engineering and Mechanical Engineering,
Georgia Institute of Technology, Atlanta, GA 30332, USA}

\date{\today}

\begin{abstract}

Fully turbulent flows are characterized by
intermittent formation of
very localized and intense velocity gradients.
These gradients
can be orders of magnitude larger than their typical value
and lead to many unique properties of turbulence.
Using direct numerical simulations of the
Navier-Stokes equations
with unprecedented small-scale
resolution,
we characterize such extreme events
over a significant range of
turbulence intensities, parameterized by the
Taylor-scale Reynolds number ($\re$).
Remarkably, we find 
the strongest velocity gradients to empirically scale as
$\tau_K^{-1} \re^{\beta}$, with
$\beta \approx 0.775 \pm 0.025$,
where $\tau_K$ is the Kolmogorov time scale
(with its inverse, $\tau_K^{-1}$, being the 
{r.m.s.} of velocity gradient fluctuations).
Additionally, we observe velocity increments across very
small distances $r \le \eta$, where $\eta$ is the Kolmogorov length scale,
to be as large as the {r.m.s.} of the velocity fluctuations.
Both observations suggest that 
the smallest length scale in the flow behaves as
$\eta \re^{-\alpha}$, with $\alpha = \beta - \nicefrac{1}{2}$,
which is at odds with predictions from existing phenomenological theories.
We find that extreme gradients are arranged in vortex tubes, 
such that strain 
conditioned on vorticity grows on average slower
than vorticity, approximately as a power law with an exponent
$\gamma < 1$, which weakly increases with $\re$.
Using scaling arguments, we get $\beta=(2-\gamma)^{-1}$, 
which suggests that $\beta$ would also slowly increase 
with $\re$. 
We conjecture that approaching
the limit of infinite $\re$, the flow is
overall smooth, with intense velocity gradients over scale $ \eta \re^{-1/2}$,
corresponding to $\beta = 1$.

\end{abstract}

\maketitle

\section{Introduction}



\begin{figure*}[t]
\begin{center}
\hspace{-1.0cm}
\includegraphics[width=1.0\textwidth]{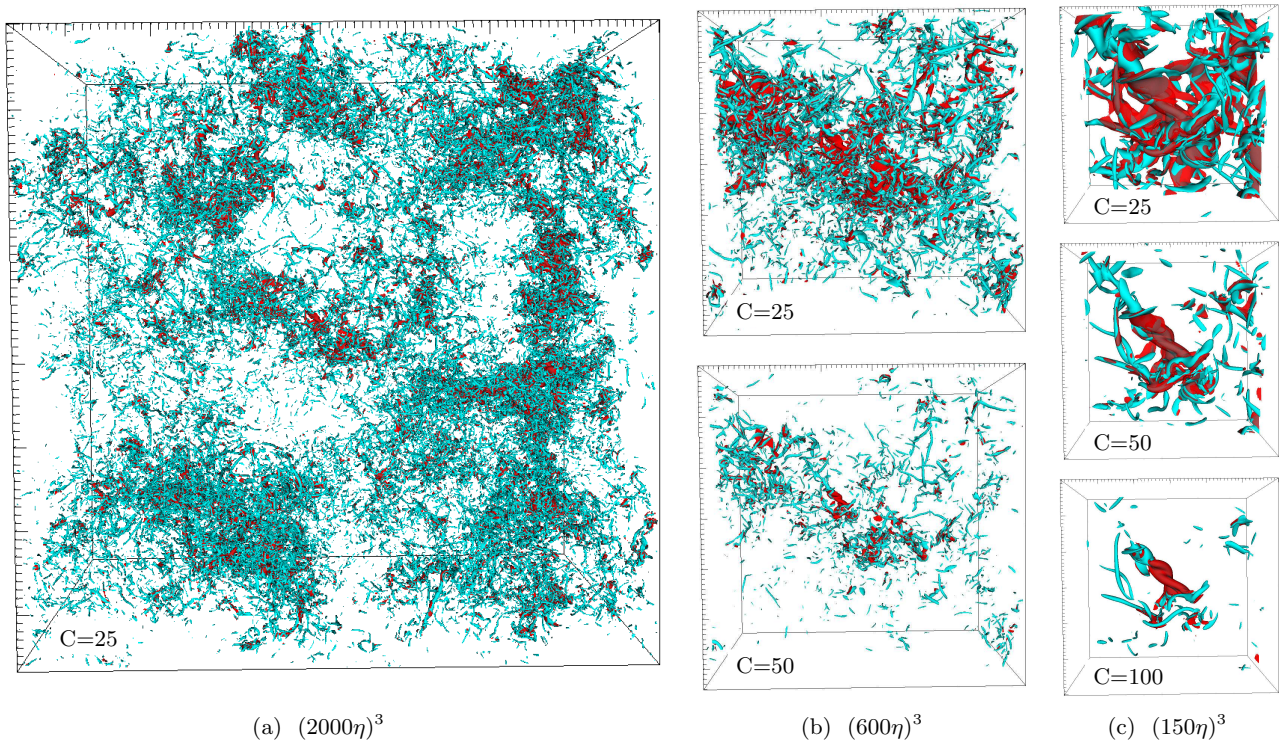}
\caption{3D-contour surfaces (in perspective view) of
enstrophy (cyan) and dissipation (red), two common measures
of the strength of the velocity gradients, normalized by their
mean values (see Section~\ref{subsec:PDF_Om_Str} for a precise definition).
The fields correspond to
a randomly chosen (but representative)
instantaneous snapshot from our numerical simulation at Taylor-scale
Reynolds number $\re=650$ on a $8192^3$ grid
or equivalently of size $(4096\eta)^3$, where $\eta$ is
the Kolmogorov length scale.
Starting from (a), we successively zoom in
and also increase the contour threshold in (b) and (c),
such that all sub-domains share the same center,
which corresponds to the strongest gradient in the snapshot. 
Approximate domain sizes (in terms of $\eta$) 
are indicated in the sub-captions, whereas
the contour thresholds C, are shown on the lower-left side of 
each panel.
The visualizations reveal the presence of numerous vortex tubes (cyan), organized
in a very heterogeneous (intermittent) structure, often
accompanied by intense strain (red),
over a wide range of scales.
More details about the center region are shown in
Fig.~\ref{fig:subcubes} in Section~\ref{sec:vis}.
}
\label{fig:fig1}
\end{center}
\end{figure*}

Quantitative studies of turbulence in incompressible
flows reveal that the averaged dissipation rate of turbulent kinetic energy,
$\langle \epsilon \rangle$, is independent of kinematic viscosity, $\nu$, 
when  $\nu \rightarrow 0$ 
or equivalently when the turbulence intensity,
i.e., the Reynolds number, is very high \cite{FalkSreeni:06, KM.2013}.
This empirical result, also known as the zeroth law of turbulence, 
implies that the amplitude of velocity gradients grows on average 
as $(\langle \epsilon \rangle/\nu)^{1/2}$.
However, the fluctuations of velocity gradients are orders 
of magnitude
larger than this average value, a phenomenon referred to as small-scale intermittency
\cite{Frisch95,Sreeni97}.
Such extreme events play a crucial role in numerous physical processes in 
both nature and engineering, e.g. 
turbulent dispersion \cite{falkovich01}, 
cloud physics \cite{shaw03},
turbulent combustion in jet engines \cite{Sreeni04,hamlington12},
and are also conjectured to be connected to 
regularity and smoothness of fluid equations \cite{leray,Fefferman}.
Hence, understanding their formation
and statistical properties is
of central importance in 
developing  a complete theory of turbulence \cite{Sreeni97}.
The complexity of the problem is apparent in Fig.~\ref{fig:fig1},
which shows the structure the velocity gradients.
The strongly intermittent nature of turbulence 
is clearly visible by the highly inhomogeneous distribution of the regions
of very intense gradients (see also \cite{KM.2013}). 
Fluid turbulence involves
a wide range of spatial scales, from approximately the system size all the way 
down to the very finest scale, corresponding to the largest gradients.
In this respect, it can be viewed as an emblematic example
for other complex dynamical systems, where such extreme events are also
observed~\cite{Solli:2007,PhysRevX.8.011017}, including the 
climate system~\cite{Rahmstorf:2011}, with its far-reaching implications.

Ever since Kolmogorov formulated and refined his seminal hypotheses 
\cite{K62},
intermittency in turbulence has
been the subject of many studies \cite{Sreeni97}. 
In particular, detailed investigations 
demonstrate
that the very large fluctuations in velocity
gradients become more extreme with increasing Reynolds number
\cite{MS91,Donzis:08}. 
While there have been theoretical proposals  to describe 
quantitatively the Reynolds number
dependence of velocity gradient fluctuations 
\cite{Paladin87,Sreeni88,Nelkin90,YS:05}, they 
have remained difficult to verify due to lack of reliable data. 
In fact, directly measuring 
the most intense fluctuations, 
experimentally or numerically,
over a reasonable range of Reynolds number is a very challenging endeavor, 
as very high spatial and temporal resolution is required to accurately resolve
such fluctuations.
As recently pointed out \cite{PK+18}, this demand can be even stricter
in numerical simulations than previously expected.
Consequently, such high resolution investigations have been so far restricted  to low
Reynolds numbers \cite{Donzis:08,Schum+07}. 

In this work, we characterize the dependence of
the extreme velocity gradients 
on the Reynolds number, and illuminate the underlying
physical processes.
To this end, we use
high resolution direct numerical simulations (DNS) 
of isotropic turbulence, based on highly accurate
Fourier pseudo-spectral methods. 
To accurately resolve the extreme gradients, all
our simulations were carried out with a
small-scale resolution at least 3-4 times higher than typical turbulence simulations,
along with appropriate temporal resolution \cite{PK+18}.
Going up to grids of $8192^3$ points, we have obtained 
results at Taylor-scale Reynolds number
($\re$) ranging from 140 to 650.

\textcolor{black}{
In order to characterize the gradients, we 
first consider the probability density functions (PDFs) 
of square of the norm of strain and vorticity,
which represent the symmetric
and skew-symmetric components of the velocity gradient 
tensor respectively.
They are analogous to dissipation rate and enstrophy 
and have the same mean values (within a prefactor) in isotropic turbulence, 
given by $1/\tau_K^2$, where 
$\tau_K = (\nu /\langle \epsilon \rangle)^{1/2}$ 
is the Kolmogorov time scale \cite{Donzis:08}. 
Consistent with previous works \cite{MS91,Yeung12}, we observe that 
the PDFs of 
these quantities, when normalized by $\tau_K$ 
exhibit tails that become broader with increasing $\re$.}
By further characterizing these PDFs,
we demonstrate that
their tails can be collapsed very well 
over the range of $\re$ covered here, when
instead normalized by the time scale:
\begin{align}
\tau_{ext} = \tau_K \times \re^{-\beta} \ , \ \  \beta>0 \ ,
\label{eq:def_tau_ext}
\end{align} 
\textcolor{black}{which implies
that the strongest gradients in the flow correspond to a time scale
$\tau_{ext}$ which increasingly decreases with respect to $\tau_K$
as $\re$ increases (and hence
the strongest gradients in the flow
grow as $\tau_K^{-1} \re^{\beta}$).}
Numerically, we find that $\beta \approx 0.775 \pm 0.025$.
The tails of the PDFs of velocity increments $\delta u_r$,
normalized by the Kolmogorov velocity scale 
$u_K (= (\nu \langle \epsilon \rangle)^{1/4} )$,
over distances $r \le \eta$
(where $\eta=(\nu^3/ \langle \epsilon \rangle)^{1/4}$ is the Kolmogorov length 
scale),
also become broader when $\re$ increases.
On the contrary, when normalized by  
the r.m.s. of velocity fluctuations 
$u^\prime$, the tails grow very slowly.
With the understanding that the most intense gradients in the flow occur with
velocity increments of order $u^\prime$ over a scale $\eta_{ext}$
we conclude that  $\eta_{ext} \sim \eta \re^{-\alpha}$,
with $\alpha=\beta-\nicefrac{1}{2}$, represents
the smallest scale in the flow.
The collapse of the tails of PDFs of $\delta u_r$, 
when normalized by either
$u_K \re^{\beta}$ or $u^\prime \re^{\alpha}$, supports these findings.
Comparisons with existing 
theoretical predictions \cite{Paladin87, YS:05}, 
point to difficulties in explaining our data.
However, these theories utilize 
the phenomenological definition that the smallest scales
in the flow 
correspond to a local Reynolds number of unity~\cite{Frisch95},
which, contrary to the numerical results of \cite{Jimenez93,Jimenez98} and also our own,
does not appear to be satisfied at the location of intense gradients,
as utilized in current work to characterize the smallest scales.


Consistent with earlier works 
\cite{Siggia:81,Jimenez93,Ishihara09},  
we find that the structures corresponding 
to the largest velocity gradients to be vortex 
tubes. We 
do not find extreme events in strain and vorticity to be 
colocated \cite{Donzis:08,Yeung15}.
Conditional averaging shows that 
intense strain is always likely to be accompanied by equally intense vorticity.
However, intense vorticity is found to be accompanied by relatively less intense strain,
with an approximate  power law dependence corresponding to exponent $\gamma<1$, which 
very slowly increases with $\re$. 
With the interpretation that $\eta_{ext}$ is the radius of most 
intense vortex tubes, we use simple scaling arguments to relate
it to the conditional strain, and thereby relate $\gamma$ to $\beta$.
This suggests that $\beta$ would also slowly increase with $\re$.
We conjecture that the limit $\beta=1$ (and $\alpha=0.5$), as predicted by some
intermittency theories, would only be realized for $\re \rightarrow \infty$.

The rest of the manuscript is organized as follows. In Section~\ref{sec:num},
we describe our numerical methods. 
Our numerical results concerning the
scaling of extreme velocity gradients are presented in 
Section~\ref{sec:results}.
The structure of regions of very intense velocity gradients is investigated
in Section~\ref{sec:vis}.
Section~\ref{sec:discuss} contains a discussion,
comparing our results with
existing theories, and then providing an
an alternative description connected to  
flow structure examined in Section~\ref{sec:vis}.
We briefly discuss the implications of our results on
future DNS and experiments in Section~\ref{sec:implic}.
Finally, we present our conclusions in Section~\ref{sec:concl}.

\section{Numerical approach and database} 
\label{sec:num}

The present work is based on DNS 
of the incompressible Navier-Stokes equations
\begin{align}
\partial \mathbf{u}/ \partial t  + ( \mathbf{u} \cdot \nabla ) \mathbf{u}  = 
- \nabla p/\rho + \nu \nabla^2 \mathbf{u} + \mathbf{f} \ ,
\label{eq:NS} 
\end{align}
where $\mathbf{u}$ is the velocity field 
(satisfying
$\nabla \cdot \mathbf{u} = 0$), $p$ is pressure,
and $\mathbf{f}$ is the forcing term used to maintain 
a stationary state \cite{EP88,DY2010}.
The equations are solved utilizing a massively parallel 
implementation of Rogallo's pseudo-spectral algorithm \cite{Rogallo}, 
whereby the aliasing errors are controlled by a combination of
truncation and phase-shifting~\cite{PattOrs71}.
\textcolor{black}{We use explicit second-order 
Runge-Kutta scheme for time integration,
with the time step $\Delta t$ subject to a constraint
for numerical stability expressed in terms of the Courant number,
$C = \frac{\Delta t}{\Delta x}  \left( ||\mathbf{u}||_1 \right)_{max}$,
where $||\cdot||_1$ represents the $L^1$-norm and 
the maximum is taken over all ($N^3$) grid points}.
The flow simulated is homogeneous and isotropic with 
periodic boundary conditions, on a cubic domain of $(2\pi)^3$
for all cases. 

As stressed earlier, appropriate numerical resolution of the
small scales is crucial to our study of extreme velocity gradients.
Spatial resolution in pseudo-spectral DNS is typically measured by the parameter
$k_{max}\eta$, 
where $k_{max}=\sqrt{2}N/3$ is the largest
wavenumber resolved 
and $\eta$ is the Kolmogorov
length scale. Equivalently,
one can use the ratio $\Delta x /\eta$ ($\approx 2.96/k_{max} \eta$),
where $\Delta x = 2\pi/N$ is the grid spacing. 
Most turbulence simulations, aimed at reaching high Reynolds number,
are in the range
$1 \le k_{max}\eta \le 2$ 
\cite{Yeung15,Ishihara16}.
However, resolution studies have shown that such a resolution
is inadequate for studying extreme events in velocity gradients
\cite{Schum+07,Donzis:08,PK+18}.
Hence, we have 
consistently used
$k_{max}\eta \approx 6$ in all the runs shown here.
Additionally, we have also used a Courant number of $0.3$,
instead of 0.6 in previous studies 
e.g. \cite{Donzis:08,Yeung12,Yeung15}, as it was recently found that 
the latter led to spurious over-prediction of the gradients~\cite{PK+18}.
Resolution studies presented in \cite{PK+18} and our own 
tests confirm that the resolution used here is adequate 
to address the questions asked in this work.

\begin{table}[h]
\centering
    \begin{tabular}{cccccc}
\hline
    $\re$   & $N^3$    & $k_{max}\eta$ & $T_E/\tau_K$ & $T/T_E$ & $N_s$  \\
\hline
    140 & $1024^3$ & 5.82 & 16.0 & 6.5 &  24 \\
    240 & $2048^3$ & 5.70 & 30.3 & 6.0 &  24 \\
    390 & $4096^3$ & 5.81 & 48.4 & 2.8 &  28 \\
    650 & $8192^3$ & 5.65 & 74.4 & 1.1 &  35 \\
\hline
    \end{tabular}
\caption{Simulation parameters for the DNS runs
used in the current work: 
the Taylor-scale Reynolds number ($\re$),
the number of grid points ($N^3$),
spatial resolution ($k_{max}\eta$), 
ratio of large-eddy turnover time ($T_E$)
to Kolmogorov time scale ($\tau_K$),
length of simulation ($T$) in stationary state
in terms of turnover time
and the number of instantaneous snapshots ($N_s$) 
used for each run to obtain the statistics.
}
\label{tab:param}
\end{table}

The database used here and the corresponding simulation parameters
are listed in Table~\ref{tab:param}.
The Taylor-scale Reynolds numbers ($\re$) considered here
are similar to those in some previous works \cite{Donzis:08,Yeung12},
but with a much higher small-scale resolution as emphasized earlier. 
\textcolor{black}{
These high resolution simulations were recently used in \cite{PK+18}.
In the present work, we simply restarted these runs
(which were already in a stationary state)
and extended them to substantially longer times to greatly improve 
statistical convergence.
We list the length of the current simulation $T$ 
in terms of the large-eddy turnover time $T_E$.}
The statistical results shown here were obtained by analyzing $N_s$ 
instantaneous snapshots for each run.
Whereas a long simulation is typically desirable for sampling
accuracy, finite resources have limited the value of $T$ for the
highest resolution runs.
Nevertheless, since our focus is on
highly intermittent velocity gradients,
one can improve sampling by simply analyzing more snapshots
for a given simulation length
as the Reynolds number increases. 
This is justified, both
by the increase of the ratio of time scales $T_E/\tau_K$ with $\re$,
and also by the increasingly smaller time scales associated
with the extreme events, as discussed in the manuscript.

\section{Scaling of extreme velocity gradients}
\label{sec:results}

\subsection{PDFs of vorticity and strain}
\label{subsec:PDF_Om_Str}

In order to study small-scale intermittency, 
in this sub-section,
we characterize 
the velocity gradient tensor by its
two quadratic invariants~\cite{Sreeni97}, namely
$\Omega = \omega_i \omega_i$, where 
$\mathbf{\omega} = \nabla \times \mathbf{u}$ is the vorticity,
and $\Sigma = 2 s_{ij} s_{ij}$, where $s_{ij} $ is
the strain rate tensor defined as 
$s_{ij} = ( \partial u_i/\partial x_j + \partial u_j/\partial x_i )/2$.
The former is the enstrophy and the latter is simply the dissipation
divided by viscosity i.e. $\Sigma = \epsilon/\nu$. 
In isotropic turbulence, as considered here,
$\langle \Omega \rangle =  \langle \Sigma \rangle = 1/\tau_K^2$,
where $\tau_K$
is the Kolmogorov time scale.


\begin{figure}[h]
\begin{center}
\includegraphics[width=0.45\textwidth]{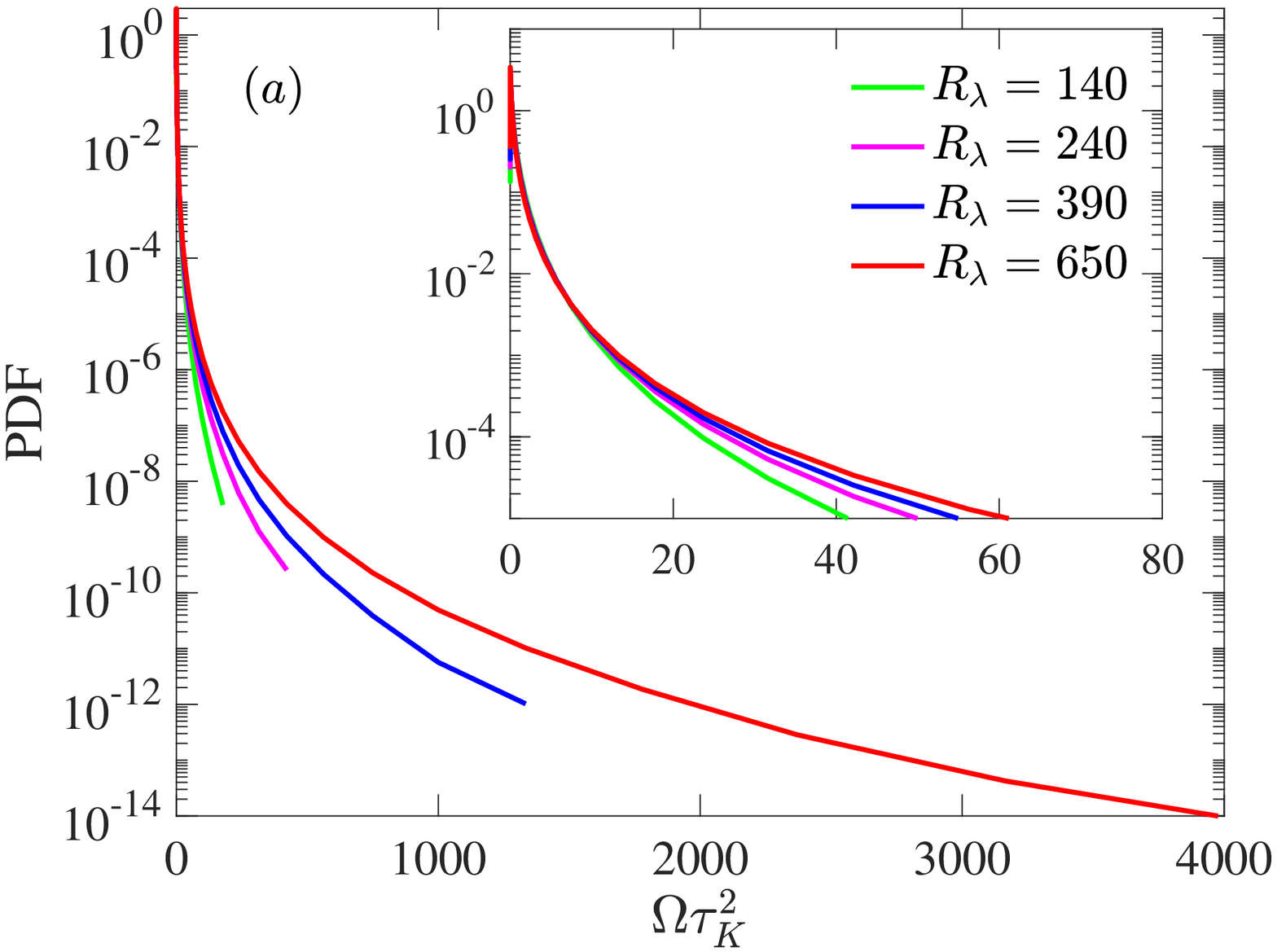} 
\includegraphics[width=0.45\textwidth]{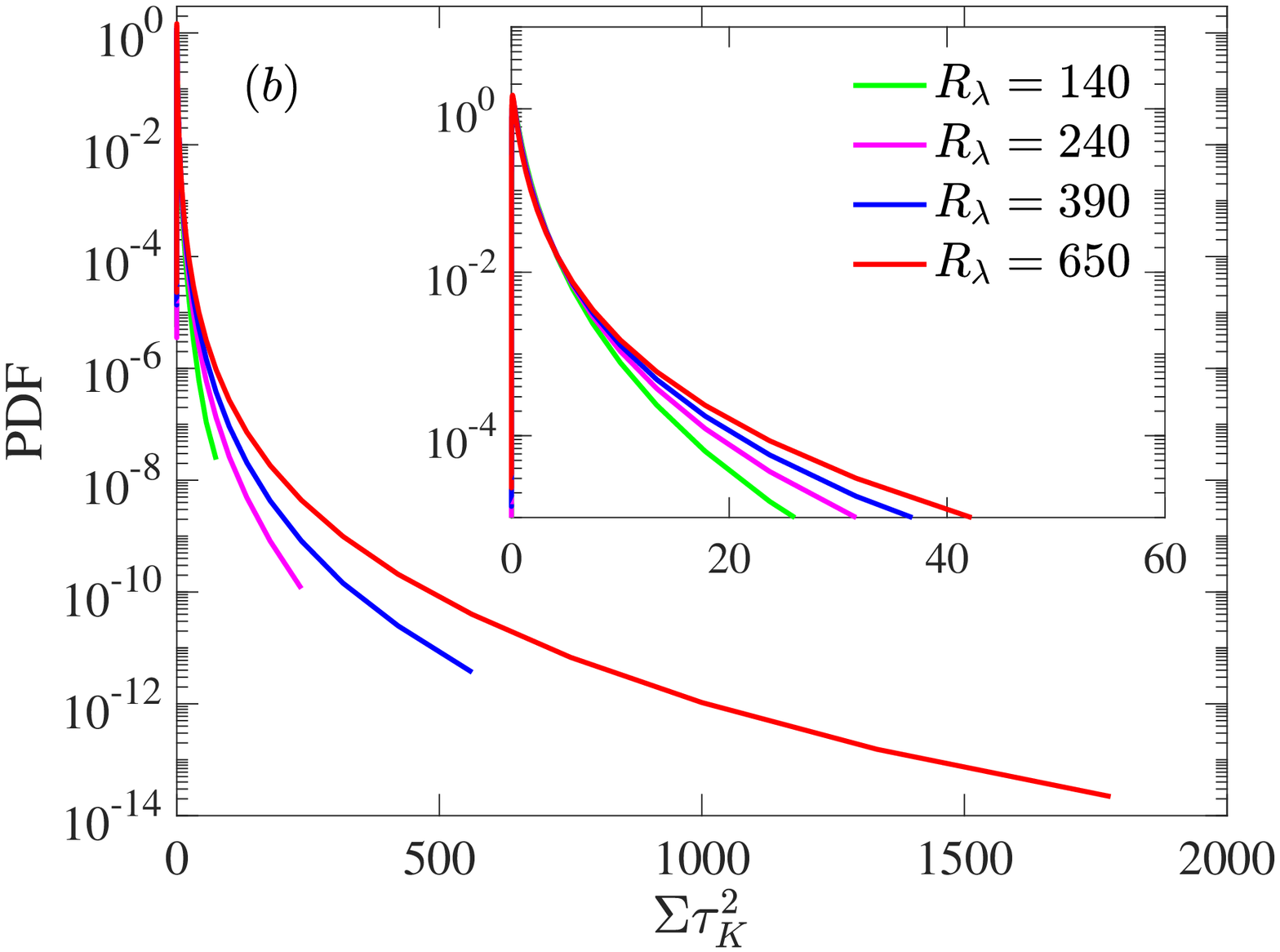}
\caption{PDFs of (a) $\Omega$ and (b) $\Sigma$, normalized
by Kolmogorov time scale $\tau_K$, for various $\re$.
\textcolor{black}{Data are shown only up to values on the x-axis where
the PDFs are statistically converged.}
The insets show zoomed in region for moderate events, revealing that
the PDFs 
\textcolor{black}{approximately superpose for 
moderate events, e.g., $\Omega\tau_K^2\lesssim 10$, $\Sigma \tau_K^2 \lesssim7$,
and thereafter start deviating systematically  as events get stronger}.
}
\label{fig:PDF_Om_eps}
\end{center}
\end{figure}

Investigating the extreme events
in $\Omega$ or $\Sigma$ amounts to focusing on the outmost parts of the
(wide) tails of their PDFs.
Issues of statistical convergence makes a precise
determination of these quantities extremely difficult.
With the available data, we estimated the statistical error 
in each bin, by looking at the sample to sample fluctuations across
various snapshots used to determine the statistics. We kept
only bins with an error less than $20\%$ compared to the mean. 
These PDFs, converged with respect to both the 
small-scale resolution \cite{PK+18} and statistical sampling, 
allowed us to determine the properties of the extreme events, as 
presented next.

Fig.~\ref{fig:PDF_Om_eps} shows the PDF of
(a) $\Omega$ and (b) $\Sigma$, normalized by their mean value, $1/\tau_K^2$,
at various Reynolds numbers.
We primarily observe that as the Reynolds number
increases, the tails of these PDFs get wider and 
extend to much higher values. 
Equivalently, the likelihood of finding a value 
of $\Omega \tau_K^2$ or $\Sigma \tau_K^2$ larger than a given large value
increases with the Reynolds number.
This is expected and consistent with previous studies.
One notices, however, that the part of the PDFs,
corresponding to events smaller than about 10 times the mean,
appear to approximately collapse for different Reynolds
numbers. This can be seen in the insets of Fig.~\ref{fig:PDF_Om_eps},
which show a zoomed in version.

\begin{figure}[h]
\begin{center}
\includegraphics[width=0.50\textwidth]{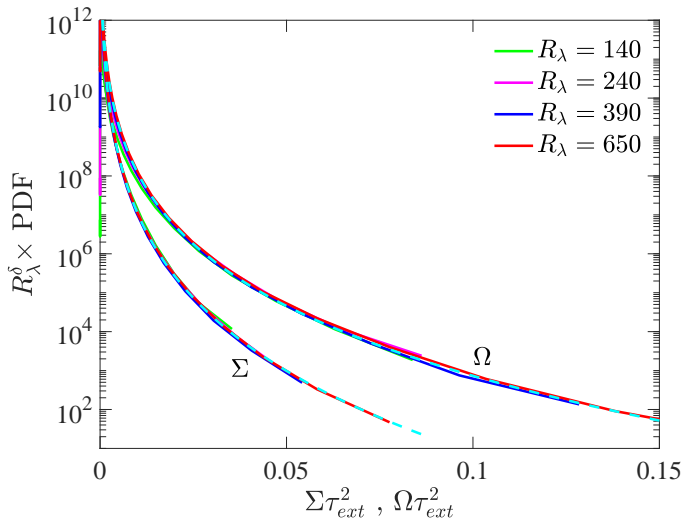}
\caption{PDFs of $\Sigma$ and $\Omega$ 
normalized by $\tau_{ext}^2$, as defined by Eq.~\eqref{eq:def_tau_ext} with
$\beta = 0.775$ and also rescaled with a factor $\re^\delta$, with $\delta\approx4.0$.
The dashed lines (cyan) show the corresponding fit by a stretched exponential
corresponding to Eq.~\eqref{eq:str_exp_fit} and Eq.~\eqref{eq:abprime}, 
with $b^\prime_{\Omega} \approx 58.0$ and 
$b^\prime_{\Sigma} \approx 46.6$. 
}
\label{fig:PDF_Om_eps_rscl}
\end{center}
\end{figure}

The existence of increasingly large fluctuations, as 
shown in Fig.~\ref{fig:PDF_Om_eps}, leads us to ask
how large are the extreme gradients and how quickly do they grow
with increasing Reynolds number.  
We propose to answer this by rescaling the PDFs. 
Namely, we use a different time scale, 
$\tau_{ext} = \tau_K \times \re^{-\beta}$ 
(see Eq.~\eqref{eq:def_tau_ext}), to rescale the extreme 
values. 
Denoting $f_{\Omega}(\Omega_e)$ and 
$f_{\Sigma}(\Sigma_e)$ as 
PDFs of $\Omega_e = \Omega \tau_{ext}^2$ and 
$\Sigma_e = \Sigma \tau_{ext}^2$ respectively,
Fig.~\ref{fig:PDF_Om_eps_rscl} shows
$\re^{\delta} f_{\Omega} (\Omega_e)$, and
$\re^{\delta} f_{\Sigma} (\Sigma_e)$. The factor
$\re^{\delta}$ provides a measure of how rare the
largest fluctuations of $\Omega_e$ or $\Sigma_e$ are, when $\re$ increases.
As shown in Fig.~\ref{fig:PDF_Om_eps_rscl},
using $\beta \approx 0.775$ and $\delta \approx 4.0$, 
the wide tails of rescaled PDFs are almost perfectly collapsed.
This indicates that while the average events in 
$\Omega$ and $\Sigma$ scale as $\tau_K^{-2}$, the most extreme events
behave like 
$\tau_K^{-2} \re^{2\beta} $.

The exponents $\beta$ and $\delta$, used in
Fig.~\ref{fig:PDF_Om_eps_rscl} to collapse the large tails of the PDFs,
can be also empirically determined 
by utilizing a functional form of the tails of PDFs of $\Omega$ and $\Sigma$. 
While theories have proposed several functional forms 
for the entire range of PDFs  \cite{benzi91,Sreeni97,yakhot06,wilczek09}, 
the stretched exponential function is known to empirically fit 
the tails of the PDFs  very accurately \cite{MS91,KSS92,zeff:2003,Donzis:08,PK+18}.
Since the tails of the PDFs in Fig.~\ref{fig:PDF_Om_eps_rscl} collapse,
we use the following 
stretched exponential functional form to mathematically
verify the value of $\beta$:
\begin{align}
f_X(x) = a \exp \left( -b x^c \right) \ ,
\label{eq:pdfexp}
\end{align}
where 
$x=\Omega \tau_K^2$ or $\Sigma \tau_K^2$
(or alternatively $\Omega/\langle \Omega \rangle$ and 
$\epsilon /\langle \epsilon \rangle$ respectively 
in the notation of \cite{Donzis:08,PK+18})
and $a$, $b$, $c$ are the fitting parameters.
Applying a change of variable 
$x_e = x \times (\tau_K/\tau_{ext})^2$, 
where $x_e=\Omega_e$ or $\Sigma_e$,
the PDF of $x_e$ becomes
\begin{align}
f_{X}(x_e) =  a \re^{2\beta} \exp \left( - b \re^{2 \beta c} {x_e}^c \right) \ .
\label{eq:str_exp_fit}
\end{align}
The collapse shown in Fig.~\ref{fig:PDF_Om_eps_rscl}
implies that 
\begin{align}
b \re^{ 2 \beta c}  = b^\prime \ , \
a \re^{ ( 2 \beta + \delta) } = a^\prime \ ,
\label{eq:abprime}
\end{align}
such that the constants $b^\prime$ and $a^\prime$ 
are independent of $\re$.
Thus, the dependence of $b^{1/c}$ as a function of $\re$, 
provides a direct access to $\beta$.

To determine the coefficients,
we simply fit 
the logarithm of the PDF to
the functional form $(\log{a} - b x^c)$ of 
Eq.~\eqref{eq:pdfexp}.
We choose the fitting window to be $x \ge 50$, 
which sufficiently excludes the region around the mean value, where 
the PDFs  appear to collapse for various $\re$ 
(as shown in insets of Fig.~\ref{fig:PDF_Om_eps}).
\textcolor{black}{We also explicitly checked by extending
the fitting range to smaller values, but found 
that the results remained virtually unchanged.}
The determination of the three parameters $a$, $b$ and $c$ then leads
to a non-linear regression.
However, since non-linear regression can be very sensitive
to the initial guess values for the fitting parameters -- especially the
value of the exponent $c$ in this particular case -- 
determining 
the parameters directly in such a manner 
can result in significant error \cite{Fletcher87}. 
On the other hand, if the value of $c$ is known beforehand, 
then a very robust fit can be obtained,
since the fitting procedure reduces to a linear regression to 
determine only $a$ and $b$.
The exact values of $a$ and $b$ would obviously substantially differ
for different values of $c$, but this would not matter if they all provide 
the same value of $\beta$ (which as shown next, is the case).

The values of the exponent $c$ in previous numerical 
studies~\cite{Donzis:08,PK+18} were found to be 
close to the range $0.23-0.25$, 
with a possible scatter within $0.19-0.29$ and no clear
dependence on Reynolds number
(e.g. see Table 4 of \cite{Donzis:08}). 
Keeping this in mind, 
we therefore fit the PDF by assuming fixed values of $c$, 
ranging from $0.19$ to $0.29$ in increments of 0.02, and 
determine the parameters $b$ and $a$ 
for PDFs $\Omega$ and $\Sigma$  for all available $\re$.
Note that a wider range for $c$ may be considered,
but this chosen range falls within the error obtained
from a naive non-linear regression and hence
for values outside the chosen range, the
quality of fit starts deteriorating. 
To provide a measure, for the chosen values of $c$,
the coefficient of determination ($R^2$) was greater than $0.995$ for every fit.
Additionally for each $c$,
the resulting values of $a$ and $b$ are always obtained with greater than 95\% confidence,
resulting in negligible error bars.  
In fact, these values are even found
to be quite insensitive
to minor variations in the fitting window, e.g., our fits
compare extremely well with those of \cite{Donzis:08}, who considered
a fitting window of $5\le x \le 100$ for $\re \le 240$.  
In this regard, we make a note that the results
of \cite{Donzis:08} can only be trusted for $\re \le 240$,
since the higher $\re$ runs were affected by resolution
issues, as reported in \cite{PK+18}.
Nevertheless, the excellent quality of fit is evident in  
Fig.~\ref{fig:PDF_Om_eps_rscl} 
(also see Fig.9 of \cite{Donzis:08} which is also in near perfect 
agreement with our fits).


\begin{figure}[h]
\begin{center}
\includegraphics[width=0.50\textwidth]{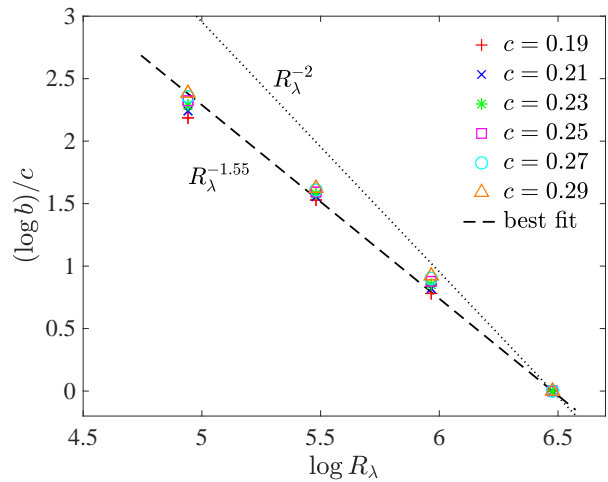}
\caption{
Plot of logarithm of $b^{1/c}$ vs $\re$
corresponding to stretched exponential fits, given by Eq.~\eqref{eq:pdfexp}, to both 
PDFs of $\Omega$ and $\Sigma$. Fits are performed for fixed values 
of $c$ ranging from 0.19 to 0.29, in increments of $\Delta c = 0.02$.
For clarity, we have divided the values of $b^{1/c}$ by its
corresponding value at $\re = 650$, so all data points exactly superpose
at $\re = 650$. 
The dashed line of slope -1.55 shows
the fit by the power law $ b^{1/c} \propto \re^{-2 \beta}$
corresponding to Eq.~\eqref{eq:abprime},
with $\beta = 0.775$.
A dotted line of slope -2, corresponding to $\beta=1$
(see discussion in Section~\ref{sec:discuss})
is also shown.
}
\label{fig:bexp}
\end{center}
\end{figure}

The dependence of the coefficient $b^{1/c}$ on $\re$ at various values of
$c$ is shown in Fig.~\ref{fig:bexp}.   
The data points correspond to curve fits for
both $\Omega$ and $\Sigma$ (thus giving two sets of points for each $c$).
For the sake of clarity, the values of $b^{1/c}$ are divided by their 
corresponding values at $\re = 650$, which imposes that all the curves
shown in Fig.~\ref{fig:bexp} pass through $1$ at $\re = 650$
(since higher $\re$ provides a larger fitting range 
and hence can be expected to be most robust). This also allows
us to directly compare the data points for every $c$ value considered.
We find that all sets of points superpose reasonably well,
and remarkably point to a similar power law in $\re$.
This collapse demonstrates that the determination of the exponent $\beta$ 
is not very sensitive to the precise
value of $c$, at least with the available data. 
However,  
weak deviations from scaling 
cannot be ruled out, especially if an even larger
range of $\re$ is considered in future. 
In fact, we will later (in Section~\ref{sec:discuss}) 
present arguments supporting a very weak growth of
$\beta$ with $\re $.

We would like to further clarify that 
ideally the best curve fits to Eq.~\eqref{eq:pdfexp}
may lead to a dependence of $c$ on the Reynolds number $\re$.
However, choosing a fixed $c$ substantially improves the quality of
curve fit and also minimizes the sensitivity to the fitting range. 
In fact, a fixed value of $c$ also helps in determining the scaling without
ambiguity, since  
if $c$ is a function of $\re$, 
the constant $b^\prime$ in Eq.~\eqref{eq:abprime} will also become a function
of $\re$, which would recursively require additional non-linear
fits to obtain $\beta$.
The minor deviations for different $c$ values at various $\re$, 
can also be possibly explained by this.
Nevertheless, the good collapse seen for such a wide range of $c$ values
provides clear evidence that the scaling proposed provides
a compelling description of our data, at least over the available range of $\re$. 
We will see that this is also 
further supported by results shown in section~\ref{subsec:PDF_inc}. 
Finally, by fitting a power law through the 
obtained data points
(marked by dashed line), 
we obtain $\beta = 0.775 \pm 0.025$ (where the error
bar takes into account the variation across different $c$ values),
which was used in scaling the PDFs in 
Fig.~\ref{fig:PDF_Om_eps_rscl}. 
The same procedure for the parameter $a$ (not shown) 
gives $\delta \approx 4.0$, with deviations of approximately 5-10\%.  
Thus, systematically characterizing the PDFs of $\Omega$ and $\Sigma$,
we are able to mathematically determine that the strongest gradients
in the flow grow as $\tau_K^{-1} \re^{\beta}$, with $\beta=0.775\pm0.025$.
We again emphasize that obtaining such a result required statistically
well converged PDFs (in turn requiring adequate spatial and temporal
resolutions) over a wide enough range of Reynolds
numbers. 

\subsection{PDFs of velocity increments}
\label{subsec:PDF_inc}

In order to further validate the scaling obtained 
in Section~\ref{subsec:PDF_Om_Str}, we next investigate the
PDFs of velocity increments. In simplified notation, velocity
increments are given as
$\delta u_r = u(x + r) - u(x)$, 
where the separation distance $r$ can be either 
in the direction of $u$ (longitudinal increments) or perpendicular to $u$
(transverse increments).
Over very small distances, the velocity differences 
essentially reduces to the velocity gradients (within 
a constant factor), aside from systematic but small errors
introduced by use of finite differencing.
Hence, we can expect velocity increments over small distances
to show the same scaling as derived earlier. 
However, to extract information about the gradients,
one still needs to ensure that $r$ is sufficiently small.
We note in this respect that the high resolution of our runs,
$k_{max} \eta \approx 6$, effectively allows us to calculate
increments over a very small distance ($r \approx \eta/2$).
A benefit of using velocity increments is that
their 1D surrogates can be also obtained and verified  using experiments \cite{EB2014}. 

\begin{figure}
\begin{center}
\includegraphics[width=0.45\textwidth]{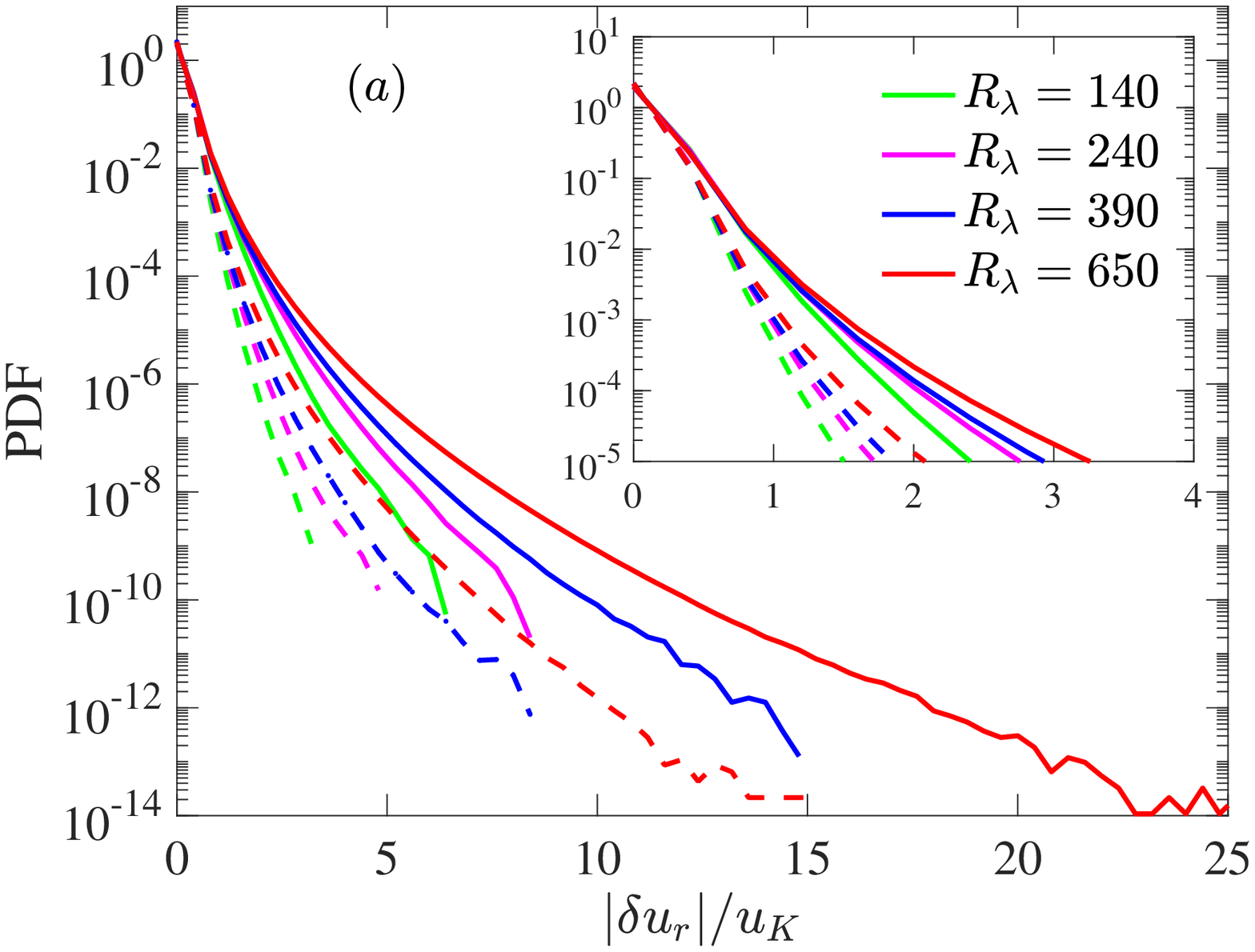} 
\includegraphics[width=0.45\textwidth]{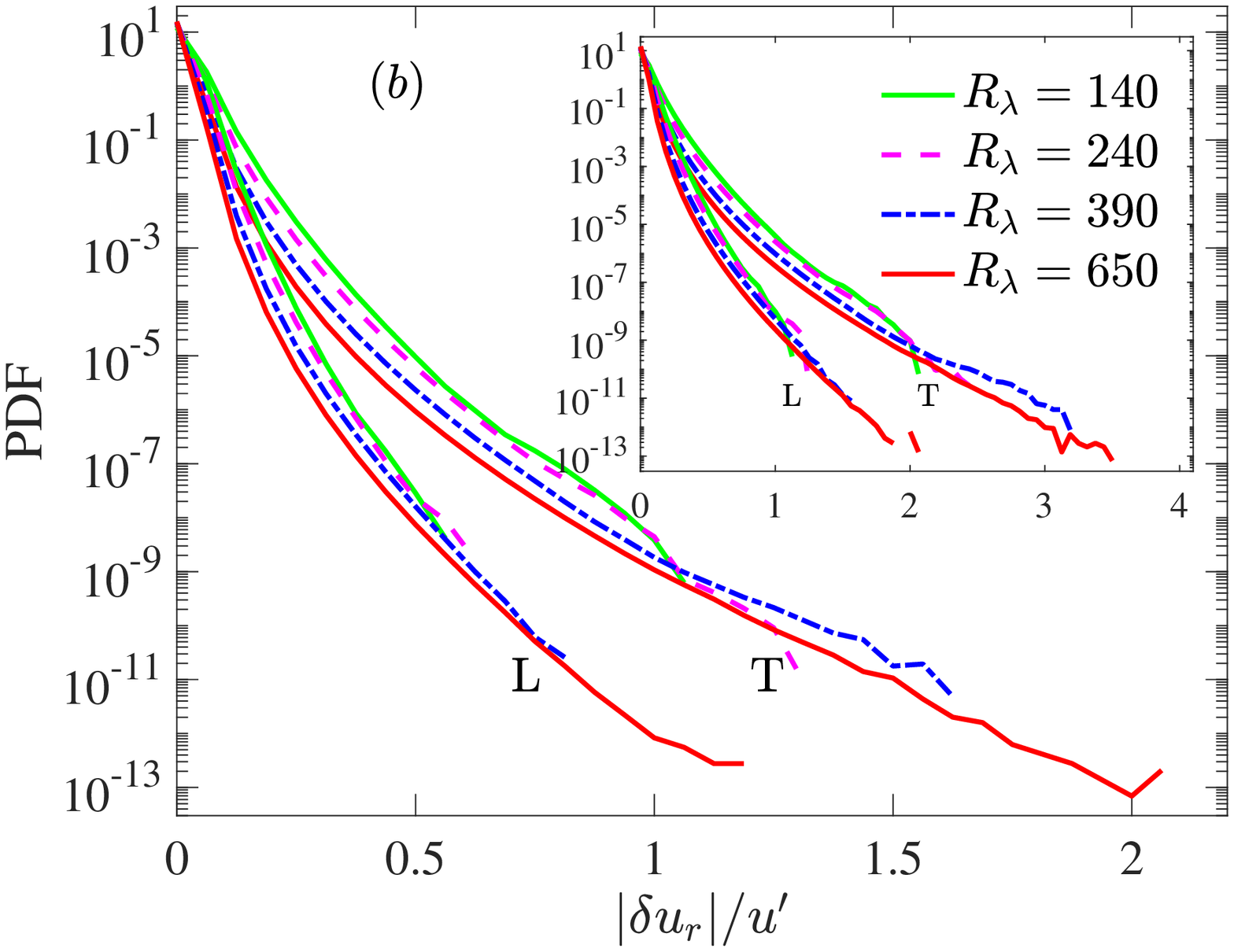} 
\includegraphics[width=0.45\textwidth]{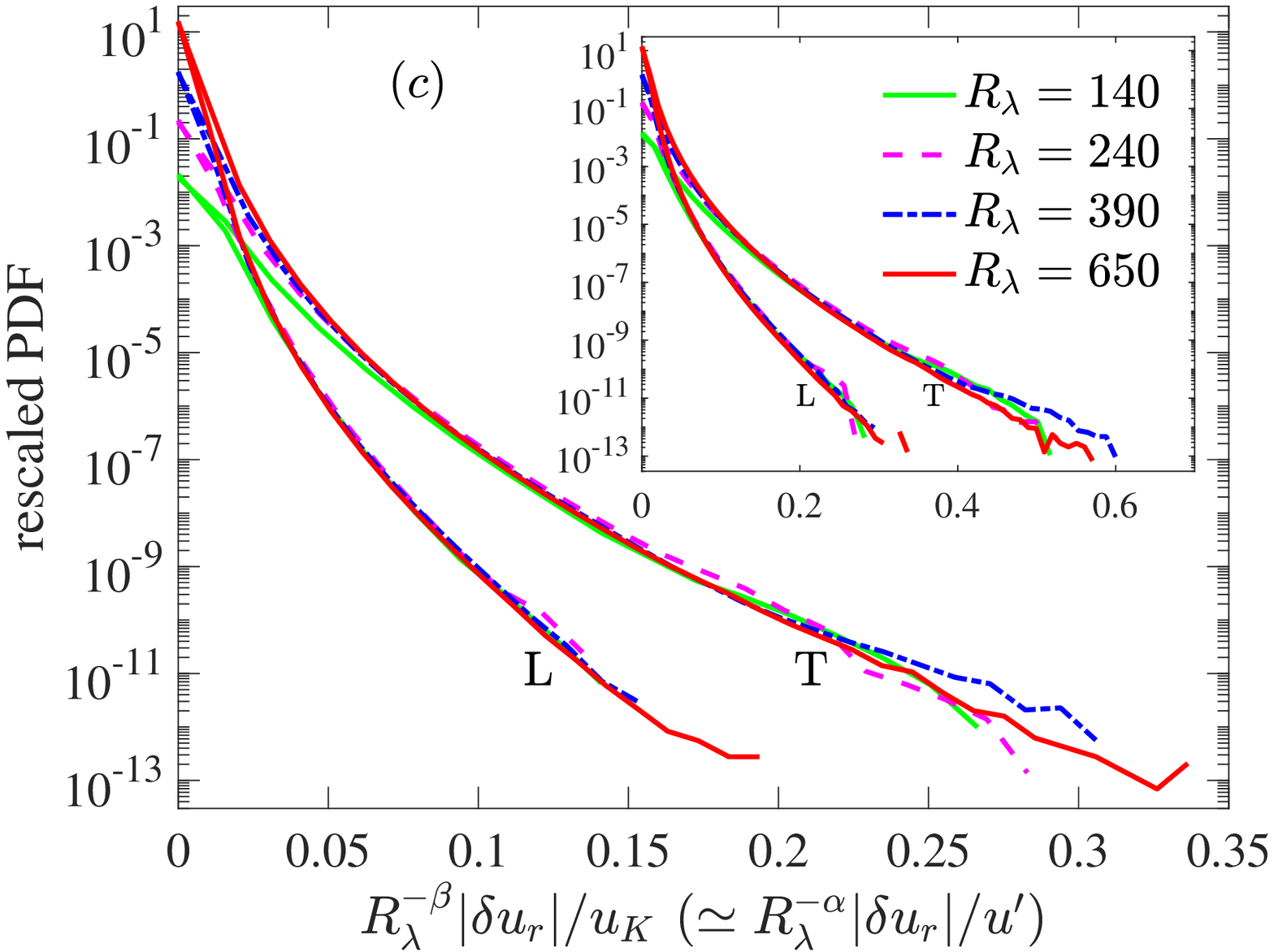} 
\end{center}
\caption{(a) PDFs of the velocity 
increments, $\delta u_r$,
normalized by $u_K$ for $r/\eta \approx 0.5$
at various $\re$. 
Solid lines for transverse
and dashed lines for longitudinal. 
The inset focuses on the
regions of weak gradients, and demonstrates that
mean events ($\delta u_r/u_K \lesssim 1$) 
collapse well.
(b) PDFs of $\delta u_r$, normalized by $u^\prime$
(r.m.s. of velocity fluctuations)
for $r/\eta\approx 0.5$ at various $\re$.
Longitudinal and transverse marked by L and T respectively.
Inset shows the same for $r/\eta\approx1$.
(c)  
Rescaled PDF of  $\re^{-\beta} \delta u_r/u_K$ 
(or equivalently $\re^{-\alpha} \delta u_r/u^\prime$),
with $\beta = 0.775$ (and $\alpha=\beta-0.5$) for $r/\eta\approx0.5$
at various $\re$.
Longitudinal and transverse marked by L and T respectively. 
The inset shows the same for $r/\eta\approx1$. 
}
\label{fig:PDF_dv_ov_ueta}
\end{figure}

Fig.~\ref{fig:PDF_dv_ov_ueta}a shows the 
PDF of $\delta u_r$, normalized by the Kolmogorov velocity scale $u_K$, 
for $r/\eta \approx 0.5$, 
at various Reynolds numbers. Both the 
longitudinal and transverse components are shown
(in dashed and solid lines respectively).
Since we are interested only in the magnitude of the 
increments, we take the absolute value of $\delta u_r$.
Consistent with the results shown in Fig.~\ref{fig:PDF_Om_eps}, the
initial part of the PDFs, corresponding to moderate events, superpose very well 
(see inset of Fig.~\ref{fig:PDF_dv_ov_ueta}a)
and as Reynolds number increases, the tails start growing.
Both the longitudinal and transverse increments show the same
behavior, with the transverse component 
being expectedly larger~\cite{Frisch95}.

To further characterize the velocity increments, we next
consider the PDFs of $\delta u_r/u^\prime$,
where $u^\prime$ is the r.m.s. of velocity fluctuations.
Fig.~\ref{fig:PDF_dv_ov_ueta}b shows the PDFs of $\delta u_r/u'$
corresponding to 
$r = \eta/2$, at various Reynolds numbers and for
both longitudinal and transverse components. 
The corresponding PDFs for $r=\eta$ are shown in the inset. 
Even at such a small separations, we observe that the 
velocity differences can be as high as $u^\prime$, 
for the entire range of Reynolds number considered.
This appears to be consistent with the observation of \cite{Jimenez93,Jimenez98}
at $\re \lesssim 170$.  
Additionally, it appears that while the probability density 
for a given $\delta u_r/u^\prime$ 
decreases with increasing $\re$, 
the extent of $\delta u_r/u^\prime$ itself slowly increases. 



Following similar ideas as in Fig.~\ref{fig:PDF_Om_eps_rscl}b,
we determine the PDFs of $\delta u_r/u_K$, rescaled by $\re^{-\beta}$,
shifted by a factor $\re^{\delta}$.
The result is shown in Fig.~\ref{fig:PDF_dv_ov_ueta}c
corresponding to $r/\eta\approx 0.5$ for both
longitudinal and transverse components. 
Note for the PDFs of $\delta u_r/u^\prime$ this corresponds
to rescaling by $\re^{-\alpha}$, with $\alpha=\beta-0.5$,
since $u^\prime/u_K \sim \re^{1/2}$ \cite{Frisch95}
(the importance of $\alpha$ is discussed later in Section~\ref{sec:discuss}).
We find that the rescaled PDFs collapse very well 
for different Reynolds numbers. 
The scatter towards the very end of the tails 
(especially for the transverse component)
can be attributed 
to lack of statistical convergence for the endmost bins.
In the inset of  Fig.~\ref{fig:PDF_dv_ov_ueta}c,
we 
repeat the exercise, but now for PDFs corresponding to $r/\eta \approx 1$. 
The superposition, although comparatively worse with respect to $r/\eta = 0.5$,
still remains very good.

\begin{figure}
\begin{center}
\includegraphics[width=0.50\textwidth]{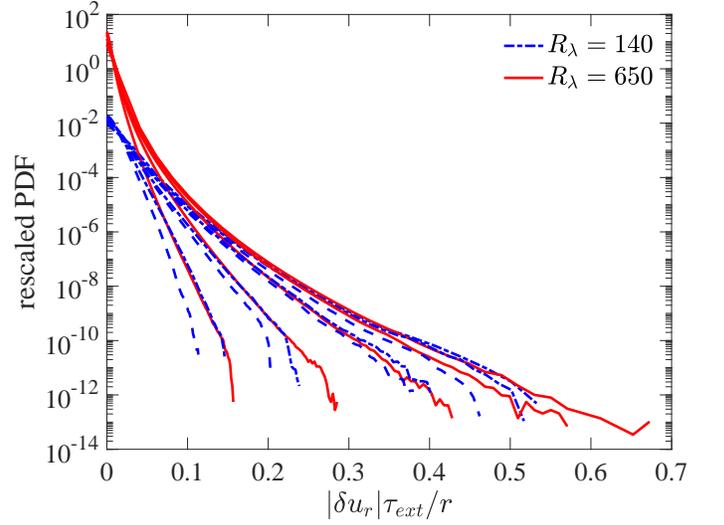}
\caption{Rescaled PDFs of the transverse velocity increments, $\delta u_r$, 
non-dimensionalized by $\tau_{ext}/r$.
Solid red lines are for $\re=650$,
showing 
$r/\eta=0.5,1,2,4,8$;
and dashed-dotted blue lines are for $\re=140$,
showing 
$r/\eta=0.5,1,2,3,3.5,6,7.5,12,16$ 
(curves for $r/\eta=2,3.5,7.5,16$ are shown with dashed lines,
see Section~\ref{subsec:new_descr} for discussion). 
Curves for increasing $r/\eta$ shift monotonically from right
to left at each $\re$.
Although not shown, the curves corresponding to
the longitudinal increments exhibit 
{\color{black} similar} behavior.
}
\label{fig:PDF_dv_ov_ueta2}
\end{center}
\end{figure}

An alternative approach
to investigate velocity increments is to consider 
the quantity $\delta u_r/r$, which 
for sufficiently small $r$, is a proxy for the
velocity gradient and thus also independent of $r$\AP{.}
Hence, it is tempting to use the same scaling based on $\tau_{ext}$ as before,
to collapse the tails of
various PDFs of $\delta u_r/r$
(as done in  Fig.~\ref{fig:PDF_dv_ov_ueta}c).
Fig.~\ref{fig:PDF_dv_ov_ueta2} shows such rescaled PDFs
for several values of $r/\eta$, at $\re=140$ and $650$.
Once again, we find that the tails of the curves for $r/\eta\approx0.5$ 
collapse very well for both $\re$. 
However at $r/\eta\approx1$, the curve for $\re=650$
starts to deviate from this collapse. 
At $r/\eta\approx2$, the curve for $\re = 650$
significantly deviates from that for $\re = 140$.
This provides yet further evidence that the resolution of
$\Delta x/\eta\approx0.5$ is adequate for both $\re$. On the other hand,  
$\Delta x/\eta\approx 1$, while adequate for $\re=140$, is
insufficient for $\re = 650$. 
Additionally, we also see that as $r/\eta$ grows, 
the deviations of the curves at $\re=650$ increase faster than those at $\re=140$. 
This also provides a hint that the smallest length scale in the flow is actually smaller than $\eta$ and 
additionally decreasing with increasing $\re$. 
This observation will be further analyzed and discussed in Section~\ref{sec:discuss}.



\section{Structure of regions of intense vorticity and strain}
\label{sec:vis}

In order to gain some understanding on the structure of
regions of intense
gradients, we first use flow visualization. 
Vorticity arranged in tube-like structures
has been repeatedly seen in DNS over a very large
range of Reynolds numbers -- from very low
at $\re \approx 45$ \cite{Siggia:81} all the way to $\re \approx 1100$ 
\cite{Ishihara09} (although the small-scale resolution in these simulations was 
limited). Whether such tubes carry the most intense
regions of velocity gradients in the flow, however, has been questioned
by a recent study~\cite{Yeung15}, which suggested that the largest 
values of $\Omega$ and $\Sigma$ appear colocated and without any coherent structure. 
It is important to note that these observations may have been affected by the numerical artefacts
documented in \cite{PK+18}. One of the motivations of the present work is 
to revisit the issue.
We again stress that the present visualizations are  based on
DNS at much higher spatial resolution than previously available. 

\begin{figure}
\begin{center}
\includegraphics[width=0.45\textwidth]{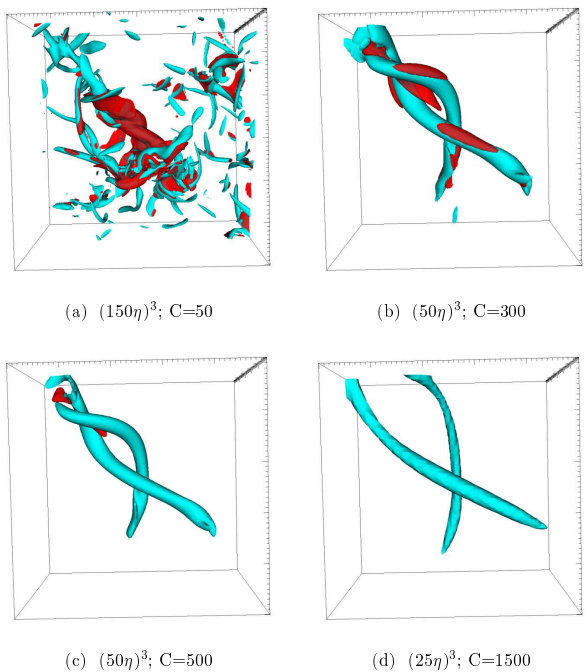}
\caption{3D-contour surfaces (in perspective view)
of $\Omega \tau_K^2$ (cyan) and
$\Sigma \tau_K^2$ (red) from the $8192^3$ simulation at $\re=650$.
The panels further zoom into the field shown in Fig.~\ref{fig:fig1}.
The middle panel of Fig.~\ref{fig:fig1}c is reproduced here in (a)
for convenience. 
The domain size in terms of the Kolmogorov length scale $\eta$,
and contour thresholds, C,  are indicated in subfigure captions.
The maximum value of $\Omega \tau_K^2$ 
is always at the center of each subcube.
No structures for $\Sigma$ are present in (d)
at the contour threshold chosen.
}
\label{fig:subcubes}
\end{center}
\end{figure}

Fig.~\ref{fig:subcubes} shows a collection of instantaneous snapshots 
from the $\re=650$ run, 
focusing on the region of most intense gradients. 
Since vorticity is in general larger than strain, 
the domains are chosen such their centers correspond to 
the maximum value of $\Omega$. 
The various panels show different contour thresholds 
(indicated as C in sub-figure caption) in
cyan for $\Omega \tau_K^2$ and in red for $\Sigma \tau_K^2$.
In the first panel (Fig.~\ref{fig:subcubes}a), a domain
of $301^3$ grid points  or $(150\eta)^3$ 
is shown with the contour threshold of 50
for both vorticity and strain. 
Structures consisting of clusters of vortex tubes,
qualitatively similar to e.g.~\cite{Ishihara09},
are readily seen.
The spacing between neighboring tubes widely varies: some vortices
are relatively isolated, others seem to be more strongly interacting
with their surrounding. Large values of strain are mostly 
located around large vortices, a phenomenon noticed many times
(see \cite{Yeung15} and references therein). 


Panel (b) zooms into the region of most intense gradients,
showing a domain of $(50\eta)^3$ with a contour threshold
of 300. 
The structure  is 
composed of two closely interacting vortex tubes, wrapped around by
intense strain.
In panels (c) and (d), 
contour levels are successively increased to 500 and 1500 respectively
and we also further zoom in to show a domain of $(25\eta)^3$ in (d).
The vortex tube structure becomes very distinct, whereas the region occupied by 
strain reduces substantially in (c) and completely disappears in (d).
Note, the largest value of $\Omega \tau_K^2$ 
is equal to about 3000 (at the center of the domain in each
panel).
In comparison, the largest value of $\Sigma \tau_K^2$ 
is about 1800, located in top left corner of domain shown in (c) --
and hence no coherent strain
region is visible in (d).  
Although not explicitly shown here, we also confirmed that the velocity
increments around the center of each panel Fig.~\ref{fig:subcubes} correspond to 
far tails of PDF of $\delta u_r$ as shown in Fig.~\ref{fig:PDF_dv_ov_ueta}, i.e.,
$\delta u_r \simeq u^\prime$.

We analyzed 
many such flow fields corresponding to
different snapshots 
and virtually all of them 
show a qualitatively similar behavior, i.e.,
as the contour thresholds are increased
tube-like vorticity structures become prominent and high-strain regions
shrink and disappear at lower values than the high-vorticity regions.
While not directly evident in Fig.~\ref{fig:subcubes},
we also find that the locations of maximum values of vorticity
and strain are typically separated by at least $10-20 \eta$ and 
never coincident, e.g. in Fig.~\ref{fig:subcubes}c, 
These observations confirm that the largest values of
$\Sigma$ are much smaller than the large values of $\Omega$ and 
the regions for large values of $\Sigma$ and $\Omega$ are not co-located.
Hence, we conclude that visualizations in \cite{Donzis:08,Yeung15} 
were also affected by
resolution issues reported in \cite{PK+18}.
We remark in this respect that new independent tests at $\re = 1300$ and
$k_{max} \eta = 3$, with a time step twice smaller than in \cite{Yeung15} -
although not shown here - confirm that the qualitative aspect of the regions of
extreme vorticity/strain are 
similar to that shown in Fig.~\ref{fig:subcubes}.

\begin{figure}
\begin{center}
\includegraphics[width=0.45\textwidth]{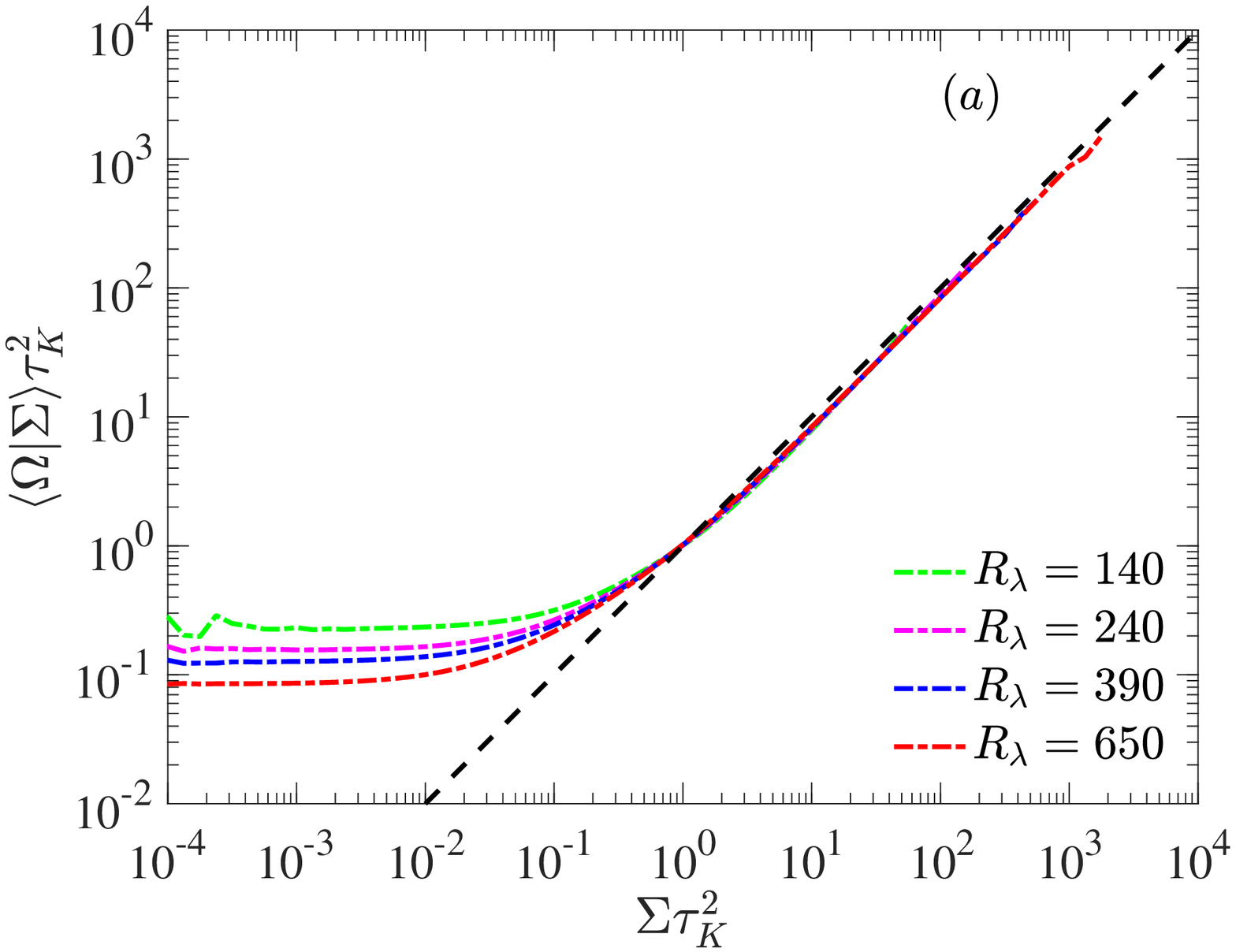} 
\includegraphics[width=0.45\textwidth]{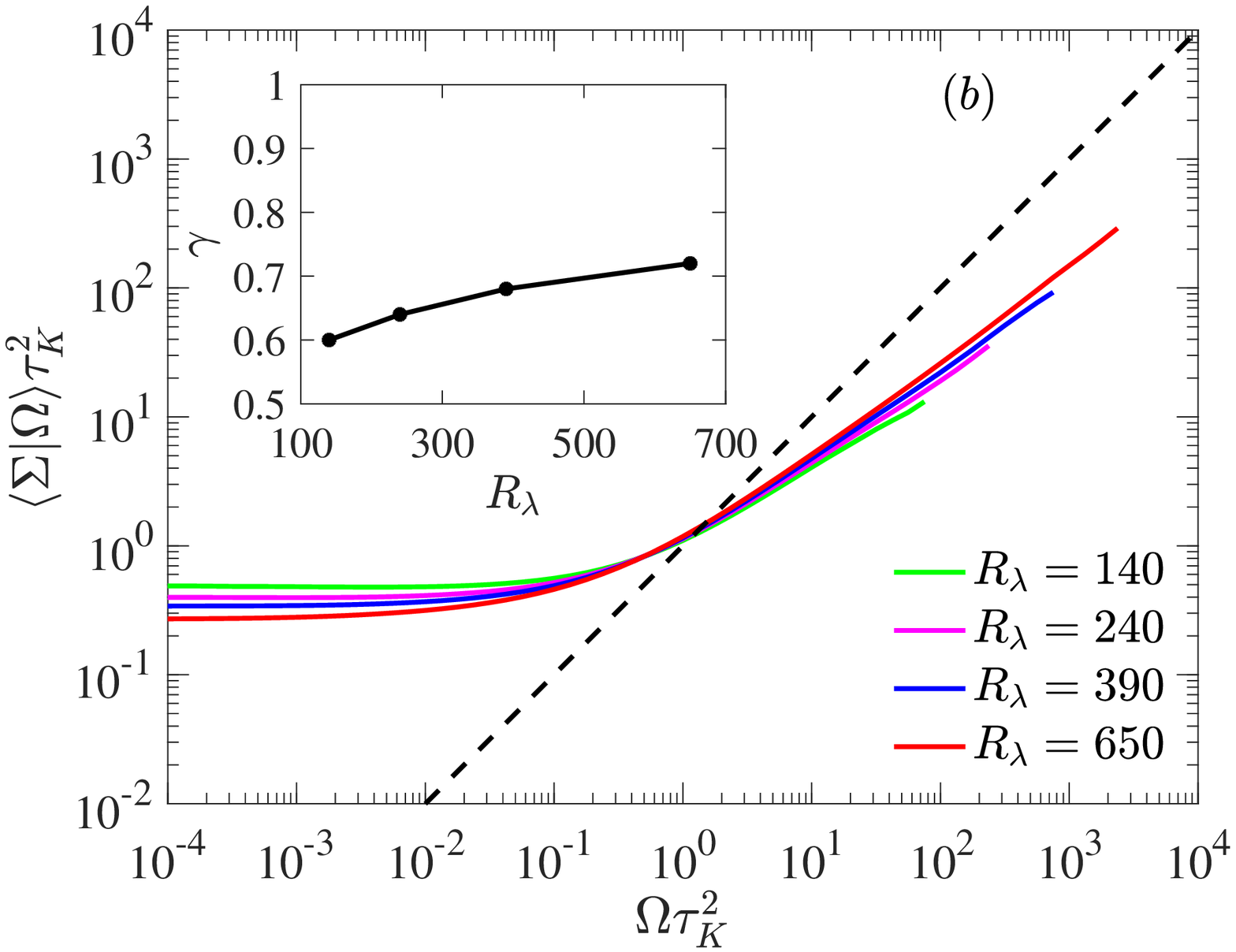}
\caption{Conditional expectations 
(a) $\langle \Omega | \Sigma \rangle$ and
(b) $\langle \Sigma | \Omega \rangle$, 
appropriately non-dimensionalized by Kolmogorov time scale $\tau_K$, 
for various $\re$. The black dashed line in both panels represents a slope of 1.
Inset in (b) shows $\gamma$ as a function of $\re$, 
for a power law $\langle \Sigma | \Omega \rangle \propto \Omega^\gamma$ 
applied in the region $\Omega\tau_K^2 \gtrsim 10$. 
}
\label{fig:cexp}
\end{center}
\end{figure}

In order to quantify the 
relation between strain and vorticity,
we next consider their conditional expectations
with respect to each other -- shown in
Fig.~\ref{fig:cexp} 
for various $\re$.
For low values of $\Omega$ or $\Sigma$, the conditional
dependencies are very weak, i.e., strain and vorticity
appear to be decorrelated. 
However, for conditional values greater than unity, i.e.,
the mean value, the conditional expectations 
clearly increase, seemingly showing a power law.
Comparison with a dashed line of slope 1 (on log-log coordinates), suggests
that 
$\langle \Omega|\Sigma \rangle  \sim \Sigma^1$. In contrast   
\begin{align}
\langle \Sigma|\Omega \rangle \tau_K^2 \sim (\Omega \tau_K^2)^{\gamma} \ , \ \ \ \gamma < 1 \ . 
\label{eq:cond_sig_om}
\end{align}
This implies that intense events in strain are always likely to be 
accompanied by equally strong events in vorticity, whereas 
the strain is comparatively weaker in very intense vortices.
This appears to be consistent with the earlier
observations of vorticity being more intermittent than
strain \cite{Siggia:81,Chen97,Donzis:08}
and ultimately concerns with
the inter-relationship of  vorticity and strain,
which is still an open question in turbulence.
Note that \cite{Donzis:08} shows a similar
plot as Fig.~\ref{fig:cexp}, however their curve for $\langle \Sigma|\Omega \rangle$ 
spuriously approaches a slope of 1 (for large $\Omega$) because of 
resolution issues~\cite{PK+18}. 

Interestingly, Fig.~\ref{fig:cexp}b also suggests that the 
exponent $\gamma$ slowly increases with $\re$.
By fitting approximate power laws, we find that $\gamma$ varies 
from $0.60-0.72$ over the range of $\re$ considered here 
(see inset of Fig.~\ref{fig:cexp}b), although
the variation appears to get weaker as $\re$ increases. 
This naturally leads to the question of what the limit of $\re \to \infty$
entails, which our data is unable to answer conclusively.
Theoretical considerations suggest that $\gamma=1$ 
in the large $\re$ limit \cite{Sreeni97,He98,Nelkin99}.
Given the very slow increasing trend of $\gamma$, it is evident that
extremely high $\re$ would be necessary to realize $\gamma=1$, if at all possible
(a simple sigmoidal or power law extrapolation 
suggests $\gamma=0.99$ would be realized for $\re\gtrsim20000$). 
Therefore, the differences between strain and vorticity
are expected to persist, even at the highest turbulence
levels on earth. 
A fundamental understanding of Eq.~\eqref{eq:cond_sig_om} from first
principles, i.e. a determination of the strain acting on a given vortex,
resulting from the tangle of vortices as shown in Fig.~\ref{fig:subcubes}
is still an open question in turbulence. As we will suggest in 
Section~\ref{subsec:new_descr}, the power law dependence on the strain 
conditioned on vorticity in fact provides a way to understand the
scaling exponent $\beta$.

\section{Theoretical considerations}
\label{sec:discuss}

In this section, we discuss the observation that the extreme velocity gradient
fluctuations scale as 
$\tau_{ext}^{-1} = \tau_K^{-1} \times \re^\beta$.
We first compare our result with existing theories,
especially the multifractal model, 
and thereafter provide a new description 
for the observed scaling which relates to the 
structure of the flow
discussed earlier in Section~\ref{sec:vis}.


In simplest terms, extreme velocity gradients result from large velocity differences 
over a very small length scale;
the largest velocity gradient in the flow can be written as
proportional to 
$\delta u_{max}/\eta_{ext}$,
where $\delta u_{max}$ is largest velocity difference
over the smallest length scale $\eta_{ext}$ \cite{Nelkin90},
which given the flow structure, 
can be physically interpreted as the radius of the smallest
vortex tube \cite{Jimenez93}.
Our notation that the largest velocity gradients scale as $\tau_{ext}^{-1}$
therefore implies 
\begin{align}
\tau_{ext} \sim \eta_{ext}/\delta u_{max} \ .
\label{eq:grad}
\end{align}
Thus, the question how large the gradients can grow entails
answering how large $\delta u$ can become over the smallest
length scale $\eta_{ext}$.
Based on earlier resolution studies \cite{Donzis:08,PK+18} and also
on the results
presented in Section~\ref{subsec:PDF_inc}, 
the smallest scale $\eta_{ext}$ can be defined by the resolution at which 
the PDFs of the gradients have converged -- which for the present
range of $\re$, gives $\eta/2 \le \eta_{ext} \lesssim \eta$.
Notice that one could formally define scales smaller than
$\eta_{ext}$, but given a smooth velocity field,
the velocity increments at such scales will simply decrease linearly 
with the scale size,
with respect to those at $\eta_{ext}$ (as also demonstrated by the
velocity increment PDFs for $r/\eta=0.5$ and $1$ at $\re=140$ in Fig.~\ref{fig:PDF_dv_ov_ueta2}). 
Thus, any length scale smaller
than $\eta_{ext}$ would show the same scaling as $\eta_{ext}$ itself 
and would be immaterial for the purpose of present study.


\subsection{Comparisons with existing theories}

It is natural to interpret our results using existing theories. 
To this end, we begin by reviewing the 
multifractal model, which provides explicit predictions concerning the smallest scales in the flow. 
In the multifractal model, as well as in some other phenomenological approaches,
a recurring concept is that of the 
fluctuating local
viscous cutoff scale, say $\eta_x$, defined such that the velocity increment over this
scale has a local Reynolds number of unity \cite{Paladin87,YS:05}:
\begin{align}
\delta  u \ \eta_x/\nu \approx 1 \ , 
\label{eq:re1}
\end{align} 
which essentially results from equating the 
viscous time scale $\eta_x^2/\nu$ to the convective time
scale $\eta_x/\delta u$.
In the multifractal framework, the velocity increment 
over a distance $r$ is given as $\delta u_r/u^\prime \sim (r/L)^h$,
where $L$ is the energy-injection scale,
and $h$ is the local
H\"older exponent within an interval $\left[h_{min},h_{max}\right]$
such that a fractal set $D(h)$ can be determined for every $h$.
Thereafter, following the derivation of \cite{Paladin87}, 
the smallest scale in the flow can readily be obtained corresponding
to the minimum  H\"older exponent
\begin{align}
\eta_{ext} \sim \eta  \re^{-\alpha} \ , \ \ \rm{where} \ \alpha = \frac{1-3h_{min}}{2(1+h_{min})}
\label{eq:eta}
\end{align}
where $\eta$ is the Kolmogorov length scale. 
It also follows
\begin{align}
\tau_{ext} &\sim  \tau_K \re^{-2\alpha} \ ,  \label{eq:tau}  \\
\delta u_{max} &\sim  u^\prime \re^{\alpha-0.5} \ , 
\label{eq:umax}
\end{align}
which implies $\beta=2\alpha$.

Earlier works have suggested that $h_{min}=0$ 
\cite{Paladin87,Nelkin90},
which gives $\alpha=0.5$, $\beta=2\alpha=1$ 
and hence $\delta u_{max} \sim u^\prime$.
The value of $\beta \approx 0.775$ derived earlier
can be obtained by using $h_{min}\approx0.06$,
and would additionally imply $\alpha = \beta/2 \approx 0.39$.
However, $h_{min}\approx 0.06$, or rather
any non-zero positive value of $h_{min}$, implies $\alpha<0.5$,
and hence suggests 
that $\delta u_{max}/u^\prime$ would 
decrease with increasing $\re$. In fact,
$h_{min}>0$ also suggests 
that the range of $\delta u_r/u^\prime$ 
decreases with $\re$ for a fixed $r/\eta$. 
Our observation in
Fig.~\ref{fig:PDF_dv_ov_ueta}b, which demonstrates that the range of 
$\delta u_r/u^\prime$
does not show any sign of decreasing with $\re$ at $r/\eta\lesssim1$ --
and rather appears to be slowly increasing, does not unambiguously support 
the decay of $\delta u_{max}/u^{\prime}$ implied by the theory. 
At the same time, 
since $\eta/2<\eta_{ext}\lesssim\eta$ over the present
range of $\re$,
the extent of PDFs 
in Fig.~\ref{fig:PDF_dv_ov_ueta}b 
implies $\delta u_{max} \gtrsim u^\prime$ for the strongest gradients. 
We notice that the probability density for $\delta u \gtrsim u^\prime$ appears
to decrease slowly with $\re$,
however, the excellent 
superposition of the PDFs in 
Fig.~\ref{fig:PDF_dv_ov_ueta}c indicates
that the decay is at best algebraic
(decreasing as $\re^{-\delta}$), and
therefore, the probability should remain
finite even as $\re \rightarrow \infty$.
The above suggests $h_{min}=0$ to allow 
$\delta u_{max} \sim u^\prime$, leading to $\beta=1$ (for $\alpha=0.5)$
as suggested by \cite{Paladin87,Nelkin90},
but at odds with the observed value of $\beta=0.775$ 
(and the corresponding $h_{min}\approx0.06$).


In our view, the inconsistency above is a result of the 
assumptions built into the extension of the multifractal theory,
originally developed to describe the inertial scales, to far dissipative 
scales. In particular, the definition in Eq.~\eqref{eq:re1}
obtained by equating the convective and dissipative time scales,
while reasonable for inertial range (where the rate of energy transfer across scales 
can be assumed to be constant), does not appear justified at 
smallest scales,
where dissipation dominates. 
This is readily observed in Fig.~\ref{fig:PDF_dv_ov_ueta}a,
where the local Reynolds number from the tails (corresponding to 
strongest gradients residing in vortex tubes) increases steadily with the $\re$
(much strongly than $\eta_{ext}$ decreases with $\re$).
In fact, earlier works based on DNS 
at $\re \lesssim 170$,
have already have shown that the local Reynolds numbers corresponding to
the vortex tubes where the intense gradients are localized
are much larger than unity,
and appear to scale differently from the multifractal prediction
\cite{Jimenez93,Jimenez98}. 
Our numerical results at significantly higher $\re$ 
(and also higher small-scale resolution) further strengthen
this conclusion
and puts into question the (phenomenological) criterion that the
smallest scale in the flow can be determined by a local Reynolds number of 
order unity and hence, also the relation $\alpha=\beta/2$.
However, an additional remark is necessary in this context. While we associate
the smallest scales of motion with the extreme gradients (which appear to reside in vortex tubes),
the theoretical constructs resulting from multifractal considerations are based on the local 
scaling, where the smallest scales simply correspond to the
minimum H\"older exponent $h_{min}$, without any
explicit connections to the flow structure. 
As a result, there is no guarantee that the scales resulting from $h_{min}$ 
actually correspond to the structures observed in 
Fig.~\ref{fig:subcubes}, calling for some caution when comparing our
results with the multifractal theory.


An alternative description, which also utilizes Eq.~\eqref{eq:re1},
is that of Yakhot and Sreenivasan \cite{YS:05}. 
In their approach, the even moments of the velocity
increment ($\delta u )^{2n}$ (or the structure functions), 
each correspond to a unique dissipative length scale 
$\eta_n$,
such that the smallest possible scale in the flow corresponds to $n\to\infty$.
Thereafter, again utilizing Eq.~\eqref{eq:re1}, $\eta_n$ can be related
to the anomalous inertial range scaling exponents of structure functions 
and $n\to\infty$ results in a similar prediction
as that of multifractal theory with $h_{min}=0$, i.e., 
$\alpha=0.5$.
However, once again, such an approach does not appear as justified 
given the lack of evidence for Eq.~\eqref{eq:re1}
to define the smallest scales.
In fact, numerical results 
of \cite{Donzis:08,Schum+07,schum07sub} all suggest that the 
smallest scales in fact grow weaker than the prediction of Yakhot-Sreenivasan
theory 
(and hence also that of multifractal theory for $h_{min}=0$).
This observation is once again reinforced by the results presented in current work.

\subsection{Alternative description in light of strain-vorticity dynamics}
\label{subsec:new_descr}

In view of the apparent shortcomings of intermittency theories  reviewed in
the previous subsection, we propose, in order to reconcile 
our observation concerning the 
very large velocity differences 
and the exponent $\beta < 1$, a different description
that directly relates to the flow structure explored in 
Section~\ref{sec:vis}. 

\textcolor{black}{Based on Fig.~\ref{fig:PDF_dv_ov_ueta}b, 
we propose that the strongest gradients correspond to 
$\delta u_{max} \sim  u^\prime$ over the smallest
scale $\eta_{ext}$. 
This is in line 
with observations of \cite{Jimenez93,Jimenez98}
and also $h_{min}=0$ as postulated by \cite{Paladin87}.}
\textcolor{black}{Note, this assumption
also essentially implies 
that the PDFs of velocity
(and hence velocity increments) are {\em bounded} \cite{leray}.}
Thereafter, substituting $\delta u_{max}\sim u^\prime$ 
and $\tau_{ext}=\tau_K\re^{-\beta}$ 
into Eq.~\eqref{eq:grad} gives  
\begin{align}
\eta_{ext}  \sim \eta \re^{-\alpha} \ , \ \ \rm{with} \ \alpha = \beta-0.5 \ ,
\label{eq:ab5}
\end{align}
where we have used $u^\prime/u_K\sim\re^{1/2}$ from classical
scaling estimate \cite{Frisch95}. 
The value $\beta=0.775\pm0.025$ found numerically
leads to $\alpha=0.275\pm0.025$, 
\textcolor{black}{which is also the value used in Fig.~\ref{fig:PDF_dv_ov_ueta}c
to collapse the PDFs of $\delta u_r/u^\prime$ for $r=\eta/2 < \eta_{ext}$.
Note, the convergence of $\delta u_r/r$ to the velocity
gradient ensures that the PDFs have a well-defined limit for
$r \le \eta_{ext}$ (and hence the PDFs at $r < \eta_{ext}$ 
can be simply obtained by linearly rescaling
the PDF at $r=\eta_{ext}$ by a factor $r/\eta_{ext}$, 
which essentially is the same as $\re^{\alpha}$ for $r=\eta/2$.
} 

This value of $\alpha\approx0.275$ can be further verified
by considering the PDFs shown in Fig.~\ref{fig:PDF_dv_ov_ueta2}.
\textcolor{black}{While for $r \le \eta_{ext}$, $\delta u_r/r$ converges to
the velocity gradient, systematic deviations
arise for $r > \eta_{ext}$.
These deviations from the gradient can be accordingly 
quantified by analyzing the higher order terms in a Taylor series 
expansion of 
$\delta u_r/r$ and can be shown to be 
approximately proportional to $r/\eta_{ext}$ \cite{Donzis:08,YS:05}.}
Since $\eta_{ext}/\eta$ decreases when $\re$ increases, at 
a given value of $r /\eta$,  
the deviations from the PDFs, especially in the tail,
from their limiting form at 
$r/\eta_{ext} \ll 1$ also increases, as clearly seen in 
Fig.~\ref{fig:PDF_dv_ov_ueta2}.
In addition, we find that the deviations of the PDF tails at 
fixed values of $r/\eta_{ext}$ to be independent of $\re$.
For $\alpha\approx0.275$, $\eta_{ext}/\eta$ decreases by approximately
$1.53$ between $\re=140$ and $650$. In contrast, taking the value of 
$\alpha$ predicted by the multifractal theory,
$\alpha=\beta/2\approx0.39$, leads to a variation of $1.82$ in the
ratio $\eta_{ext}/\eta$.
The $r/\eta$ values at $\re=140$ shown in Fig.~\ref{fig:PDF_dv_ov_ueta2} 
are chosen, as close as possible, 
within these factors (of $1.53$ and $1.82)$,
compared to the $r/\eta$ values for $\re=650$
(the curves corresponding to 1.53 are shown in dashed-dotted lines,
whereas curves corresponding to 1.82 are shown in dashed lines).
As visible, the PDFs corresponding to the factor of $1.53$
between the two $\re$ cases
collapse remarkably well (especially as $r/\eta$ increases), 
hence providing an alternative means to verify
$\alpha\approx0.275$. 

Interestingly, alongside some 
scaling arguments to evaluate $\alpha$,
Eq.~\eqref{eq:ab5} 
leads to two different limits
(which incidentally also correspond to
previously reported cases in literature).
The first limit corresponds to simply assuming that
the smallest length scale in the flow is  
the Kolmogorov length scale, i.e., $\eta_{ext}=\eta$.
Using this, we get $\alpha=0$ and  $\beta=0.5$, and hence 
\begin{align}
\tau_{ext} \sim  \tau_K \ \re^{-1/2} \ .
\label{eq:b12}
\end{align} 
This result was derived in
\cite{Jimenez93,Jimenez98}, based on the analysis of DNS data at
relatively low Reynolds numbers 
($\re \lesssim 170$), which the present work greatly improves upon. 
The second limit consists in taking into account 
the extreme fluctuations of the velocity gradients.
Physically, the smallest scale in a flow can be thought to result from a balance 
between
viscosity $\nu$ and strain $\Sigma$ ($=2s_{ij}s_{ij}$, as defined earlier), which leads to the
expression of the length scale: $\eta_{ext} \simeq (\nu^2/\Sigma)^{1/4}$, 
familiar in a number of contexts \cite{Burgers48}.
Assuming $\eta_{ext} = \eta$, which leads to Eq.~\eqref{eq:b12}, amounts
to a mean field approximation, consisting in replacing the strain by its
averaged value (as $\eta$ is calculated from the mean dissipation). 
Taking into account the large fluctuations of $\Sigma$
results in $\eta_{ext}$ being smaller than $\eta$ \cite{Sreeni88}. 
In this regard, the second limit can be simply derived by 
evaluating $\eta_{ext}$ using the maximum value of strain ($\Sigma_{max}$),
i.e., $\eta_{ext}=(\nu^2/\Sigma_{max})^{1/4}$. 
Thereafter, using $\Sigma_{max}\sim\tau_{ext}^{-2}$ 
and $\delta u_{max}\sim u^\prime$ based on earlier results,
it follows from Eq.~\eqref{eq:grad} 
\begin{align}
\tau_{ext} \sim \frac{\nu}{u'^2} = \frac{\nu}{u_K^2} \frac{u_K^2}{u'^2}
\sim \tau_K \re^{-1} \ ,
\label{eq:tau_ext_loc}
\end{align} 
where we have used $\nu/u_K^2 = \tau_K$ and $u^\prime/u_K \sim \re^{1/2}$.
This implies
$\beta = 1$ and $\alpha=0.5$ from Eq.~\eqref{eq:ab5},
as also predicted by intermittency
models discussed earlier \cite{Paladin87,Nelkin90,YS:05}.
\textcolor{black}{However, this is not completely surprising, as defining
$\eta_{ext}$ based on $\Sigma_{max}$ with $\Sigma_{max}\sim \nu\tau_{ext}^{-2}$ 
also leads to $\beta=2\alpha$, which is essentially the multifractal prediction,
and in conjunction with
$\beta=\alpha+0.5$ derived in Eq.~\eqref{eq:ab5} gives $\beta=1$ (and $\alpha=0.5$). 
Additionally for this scenario, the local Reynolds number,
which can be written as $\eta_{ext}^2\nu^{-1} \tau_{ext}^{-1}$ 
using Eq.~\eqref{eq:grad}, comes out to be constant
as inherently assumed in intermittency theories discussed earlier.
}

The numerically observed value of $\beta\approx 0.775$
lies between $\beta=0.5$ and $1$,
which suggests that $\eta_{ext}$ results from a strain,
intermediate between the two limits considered before.
In fact, 
this is precisely what we observed in Section~\ref{sec:vis}.
As noted earlier, 
Eq.~\eqref{eq:cond_sig_om} (and Fig.~\ref{fig:cexp}b) suggests that the 
strain acting on a very intense vortex tube is significantly
weaker than naively expected by postulating $\Sigma \propto \Omega$. 
A simplified estimate consists in
substituting $\Sigma_{max}$ in the argument leading to 
Eq.~\eqref{eq:tau_ext_loc}
by $\tau_K^{-2} (\tau_K^2 \Omega_{max})^\gamma$, as suggested by 
Eq.~\eqref{eq:cond_sig_om}. Thereafter, we get 
\begin{align}
\tau_{ext} \sim \tau_K \re^{-\beta} \ , \ \ \rm{with} \  \beta=\frac{1}{2-\gamma} \ .
\label{eq:rel_bet_gam}
\end{align}
The limits of $\beta=0.5$ and $1$ 
correspond to $\gamma=0$ and 1 respectively.
In view of the weak dependence of $\gamma$ shown in the inset of 
Fig.~\ref{fig:cexp}b, Eq.~\eqref{eq:rel_bet_gam} suggests a
dependence of $\beta$ on $\re$. The values of $\gamma $ observed
over the range of $\re$ studied here, $ 0.60 \lesssim \gamma \lesssim 0.72$,
implies a variation of $\beta$ in the range: $0.72 \lesssim \beta \lesssim 0.78$,
which is quantitatively consistent with $\beta \approx 0.775$ 
determined empirically in Section~\ref{sec:results}. The weak variation of
$\beta$ implied by Eq.~\eqref{eq:rel_bet_gam}
may also explain the slight
deviations from scaling seen for $\re=140$ in Fig.~\ref{fig:bexp}. 
In fact in Fig.~\ref{fig:bexp}, considering only data points at $\re=140$ and 240, 
the slope corresponds to $\beta\approx0.73$, which appears to be remarkably consistent with
that obtained
from $\gamma$ for these $\re$. 
On the other hand, the interesting possibility that 
$\gamma \rightarrow 1$ when $\re \rightarrow \infty$ 
would then suggest, in view of Eq.~\eqref{eq:rel_bet_gam}, that 
$\beta \rightarrow 1$, as originally expected by some theories 
(albeit corresponding to a constant local Reynolds number 
much larger than unity).
However, the very slow variation of $\gamma$ shown the inset of
Fig.~\ref{fig:cexp}b would indicate that $\beta = 1$ would be attained 
at extremely large values of $\re$, likely larger than 
practically relevant.
In this regard, the scope of existing predictions
in understanding
finite Reynolds number scaling appears to be severely limited. 


Whereas the prediction of the exponent $\gamma$ and its dependence on 
$\re$ is a very challenging task, we briefly note that
the cascade model of She and Leveque \cite{SL94}
presents a similar idea, though with shortcomings.
The model postulates that the locally 
averaged dissipation field 
$\epsilon_r$ at a scale $r$
and the corresponding moment ratios: 
$\epsilon_r^{(p)} = \langle \epsilon_r^{p+1} \rangle / \langle \epsilon_r^{p} \rangle$,
are related to the hierarchy of complex structures in the flow.
The most singular structures correspond to $\epsilon_r^{(\infty)}$,
which in the phenomenology of \cite{SL94} obeys the power law dependence:
$\epsilon_r^{(\infty)} \simeq \langle \epsilon \rangle  \left(L/r \right)^{\mu}$,
with $\mu=2/3$. 
While their original arguments were postulated for inertial scales,
if one were to extend the cascading process down 
to smallest scale, i.e., 
$r=\eta_{ext}=\left(\nu^3/\epsilon_r^{(\infty)}\right)^{1/4}$, 
it leads to $\alpha= (3\mu)/{(8-2\mu)}$.
Using $\mu=2/3$ as proposed by She-Leveque then gives $\alpha=0.3$, which 
is close to our current prediction of $\alpha=0.275\pm0.025$. 
However, they also suggest $h_{min}=1/9$  
within the multifractal formalism, which using Eqs.~\eqref{eq:eta}--\eqref{eq:umax},
gives $\alpha=0.3$, $\beta=0.6$ and $\delta u_{max} \sim u^\prime \re^{-0.2}$,
which are clearly inconsistent with our data.
Ultimately, phenomenological descriptions (as those of \cite{SL94}) are at best weakly connected
to flow structures, and typically assume a constant value of the exponents
such as
$\beta$ and $\alpha$, thus ignoring any possible dependence on $\re$,
as suggested from Eq.~\eqref{eq:rel_bet_gam} and Fig.~\ref{fig:cexp}.
Hence, it appears that the closeness of $\alpha$ between the She-Leveque model 
and our current result is only fortuitous.

In conclusion, our analysis of the most intense vortex structures 
observed in the flow relates the exponent $\beta$ with
the properties of the strain acting on vortices, and in particular with
the exponent $\gamma$ defined by Eq.~\eqref{eq:cond_sig_om}. The weak
variation of $\gamma$ with $\re$, see Fig.~\ref{fig:cexp}, implies that 
$\beta$ should increase with $\re$.
A very natural conjecture is that the symmetry between strain and 
vorticity, clearly broken at finite $\re$, will 
be restored as $\re \rightarrow \infty$, and that the exponents $\gamma$
and $\beta$ both tend to $1$, corresponding to earlier
predictions~\cite{Paladin87,YS:05}. 
Understanding the
$\re$-dependence of the strain acting on intense vortex tubes appears as
an essential question in this regard, that deserves renewed theoretical attention.


\section{Implications for simulations and experiments}
\label{sec:implic}

The identification of the smallest scale $\eta_{ext}$, characteristic of
the largest velocity gradients in the flow, which decreases faster than
$\eta$ when $\re$ increases, has some obvious consequences for the resolution
constraints required in both DNS and experiments.
In DNS, it is typical for most studies based to be performed with
a $k_{max}\eta$ or $\Delta x/\eta$ held constant across the range of $\re$
simulated (e.g. see \cite{Ishihara16,BSY.2015}).
On the other hand, in experiments, the resolution,
determined by the probe size or the data acquisition frequency, 
often gets worse
as $\re$ increases \cite{EB2014}.
The present results, however, show
that in studies focused on intermittency,
one must continuously improve $\Delta x/\eta$ as $\re$ is increased
to adequately resolve the smallest scales, i.e., 
$\Delta x/\eta_{ext}$ should be held constant across various
simulations.
In fact, this suggestion was also put forward by \cite{YS:05},
though their criterion was stricter than the present
numerical results suggest. 

The simulations presented here suggest, based on PDFs of various
components of the velocity gradient tensor,
that a resolution of $\Delta x/\eta \approx 1$ is sufficient for $\re=240$,
but barely insufficient for $\re=390$
(this is also evident from Fig.~\ref{fig:PDF_dv_ov_ueta2}). An earlier resolution
study at $\re\le240$ \cite{Donzis:08}, also supports this.
In fact, in \cite{Donzis:08}, the authors also explored a practical approach
to determine the necessary resolution based on calculating the error
between $p$-th order structure function and its analytic behavior
for small distances (obtained from Taylor series expansion),
such that the result is a function of $p$ and $\re$.
However, such expressions are limited to small values of $p$,
since the derivation retained only a small number of terms in
the Taylor expansion and hence also cannot be generalized
to estimate $\eta_{ext}$. 

Nevertheless, based on previous and current results,
it follows empirically  
that $\Delta x/\eta$ to accurately resolve
the velocity gradients should be
\begin{align}
\Delta x /\eta \approx \left(\re/\re^\ast \right)^{-\alpha} \ , 
\label{eq:res_const}
\end{align}
where $\re^\ast\approx300$ is
the reference Taylor-scale Reynolds number, at which $\eta_{ext} \approx \eta$.
The above relation provides a practical resolution criteria for future simulations
at even larger problem sizes than considered here. 
While $\alpha\approx0.275$ for the current range of $\re$, we anticipate 
newer simulations at higher $\re$ would progressively update $\alpha$ and 
also quantify its dependence on $\re$  
(though given the slow growth of $\alpha$ with $\re$, a very substantial range
of $\re$ might be required).
The constraint provided by Eq.~\eqref{eq:res_const} should also apply to 
experimental investigations,
which are currently capable of providing data at much higher $\re$ compared to DNS
\cite{EB2014,SHREK}.  A simple estimate suggests that $\eta/\eta_{ext} \gtrsim 3$ 
corresponding to these laboratory experiments at $\re \approx 6000-10000$.
While this correction is unlikely to affect the 
dynamics in wind tunnel experiments \cite{EB2014}, it might enhance
the quantum effects in liquid-He experiments at \cite{SHREK}. 
However, more quantitative studies, even at relatively lower $\re$, 
would be useful, 
since currently resolving
even $\eta$ in such high $\re$ experiments
is an outstanding technical challenge.

\section{Conclusions}
\label{sec:concl}

Using very well-resolved DNS 
of isotropic turbulence, 
both in space {\em and} time,
at Taylor-scale
Reynolds number $\re$ ranging from 140 to 650,
we have characterized the extreme fluctuations of the velocity gradients.
In particular, we focused 
on the square of vorticity, $\Omega=\omega_i\omega_i$,
and strain, $\Sigma=2s_{ij}s_{ij}$
(synonymous with enstrophy and dissipation),
which have the same mean value (equal to $1/\tau_K^2$). 
Whereas the PDFs of
$\Omega \tau_K^2$ 
and $\Sigma \tau_K^2$ superpose well around their mean values,
the extents of their tails strongly grows
as $\re$ increases. 
We find that these tails can be empirically collapsed by using a smaller
time scale
$\tau_{ext}$, defined as 
$\tau_{ext} = \tau_K \times \re^{-\beta}$,
implying that the extreme 
velocity gradients
in the flow grow as $\tau_{ext}^{-1}$.
The numerical results 
indicate that 
$\beta \approx 0.775 \pm 0.0025$. 

The above result is further validated by analyzing the PDFs of velocity
increments $\delta u_r$ at distances $r$ equal to or less than 
the Kolmogorov length scale $\eta$.  
Our results show that $\delta u_r$ 
can be as large as the velocity r.m.s. $u^\prime$, and slowly
increases with $\re$.
The excellent superposition of the rescaled PDFs of 
$\re^{-\alpha} \delta u_r/u^\prime$,
with $\alpha = \beta - \nicefrac{1}{2}$
over the range of $\re$ covered in this study,
suggests that the largest velocity gradients consist of velocity increments
of approximately  
$u^\prime$ over a size $\eta_{ext} \sim \eta \ \re^{-\alpha}$.
The existence of scales smaller than $\eta$ is consistent with 
previous ideas, although 
our results do not quantitatively support the existing phenomenological
theories.
In particular, the assumption that extreme events correspond
to a local Reynolds number of unity does not appear to be
justified
in vortex tubes, where extreme gradients are found to reside
-- as revealed by flow visualizations in 
Figs.~\ref{fig:fig1} and \ref{fig:subcubes} and also consistent with 
previous studies \cite{Jimenez93,Ishihara09}.

Further analysis of flow structures around intense gradients
in Fig.~\ref{fig:subcubes} reveals that 
vorticity and strain
are generally {\em not} spatially colocated, as suggested in some earlier 
studies \cite{Donzis:08,Yeung15}.
Conditional averaging shows that strain acting on intense
vorticity, is on average weaker than the vorticity,
and shows an approximate power law behavior given by Eq.~\eqref{eq:cond_sig_om}.
It is important to note that the structures shown in 
Fig.~\ref{fig:subcubes} taken in isolation do not lead to a strong vorticity
amplification. This suggests that the stretching,
necessary to create the very 
intense velocity gradients, in a representation of the
velocity field in terms of a Biot-Savart equation~\cite{PumSig87},
could originate from a non-local mechanism.
The observation that the strain acting on intense vortices
is significantly weaker than corresponding vorticity 
(as reflected in exponent $\gamma<1$ in Eq.~\eqref{eq:cond_sig_om})
can be viewed as 
a consequence of this non-locality. 
Using scaling analysis, we are able to quantitatively relate
$\beta$ with the exponent $\gamma$. The weak increase in $\gamma$ with
$\re$ suggests the same for $\beta$.
This leaves open the possibility that
$\beta$ could asymptote to  $1$ 
(and $\alpha$ to $0.5$), in the limit of $\re \rightarrow \infty$ --
a simple extrapolation of our data suggests that 
$\beta\ge0.99$ would require $\re \gtrsim 20,000$. 
However, such high Reynolds numbers might not be 
feasible experimentally or numerically.
To conclude, explaining the scalings discussed here, 
especially in the light of $\gamma$, remains an outstanding theoretical
challenge. Much remains to be learned by analyzing
well-resolved
data from even higher Reynolds numbers than considered here, from both 
DNS and experiments.

\section*{Acknowledgments}

We gratefully acknowledge the Gauss Centre for Supercomputing e.V. 
(www.gauss-centre.eu) for funding this project by providing computing time on the 
GCS supercomputer JUQUEEN at J\"ulich Supercomputing Centre (JSC),
where the simulations reported in this paper were primarily performed.
This work was also partly supported by the advanced supercomputing resources
provided to author P.K.Y. by 
the National Center for Supercomputing
Applications (NCSA) at  the University of
Illinois at Urbana-Champaign and  the Texas Advanced
Computation Center (TACC) at the University of Texas at Austin
which provided access to supercomputers Blue Waters and Stampede2 respectively.
D.B., A.P. and E.B. acknowledge support from 
EuHIT--European High-performance Infrastructure in Turbulence, 
which is funded by the European Commission Framework Program 7 (Grant No. 312778).
P.K.Y. was supported 
by National Science Foundation (NSF) Grants ACI-1036170 and 1640771
under the Petascale Resource Allocations program.
We thank K.R. Sreenivasan for helpful comments on
an earlier draft of the manuscript.



\end{document}